\documentclass[nofootinbib,aps,11pt]{revtex4}
\usepackage{graphicx}
\usepackage{cancel}
\usepackage{amssymb}
\usepackage{textcomp}
\usepackage{amsmath}
\usepackage{bm}
\usepackage{times}
\usepackage{epsfig}
\usepackage{color}

\begin{document}

\title{\Large RGE Effects on Neutrino Masses in Partial Split Supersymmetry}

\author{Fabian Cadiz}
\email{fabian.cadiz@polytechnique.edu}
\affiliation{Physique de la mati\`ere condens\'ee, Ecole Polytechnique, CNRS, 91128 Palaiseau, France}

\author{Marco Aurelio D\'\i az}
\affiliation{
{\small Departamento de F\'\i sica, Pontificia Universidad Cat\'olica 
de Chile, Avenida Vicu\~na Mackenna 4860, Santiago, Chile 
}}
%%%%%%%%%%%%%%%%%%%%%%%%%%%%%%%%%%%%%%%%%%%%%%%%%%%%%%%%%%%%%%%%%%%%
\begin{abstract}

We show that the running of the Higgs-gaugino-higgsino couplings present
in Partial Split Supersymmetry can severely affect the neutrino masses
generated through Bilinear R-parity Violation. We find a working scenario
where the predicted neutrino observables satisfy the experimental 
constraints when the running is neglected. After including the running,
we show that already with a split supersymmetric scale of $10^4$ GeV the 
atmospheric mass leaves the allowed experimental window, and that the 
solar mass leaves it even earlier, with a split supersymmetric scale of
$10^3$ GeV. This shows that the correct prediction of neutrino 
observables in these models necessitates the inclusion of the running
of these couplings.

\end{abstract}
%%%%%%%%%%%%%%%%%%%%%%%%%%%%%%%%%%%%%%%%%%%%%%%%%%%%%%%%%%%%%%%%%%%%
\date{\today}
\maketitle
%%%%%%%%%%%%%%%%%%%%%%%%%%%%%%%%%%%%%%%%%%%%%%%%%%%%%%%%%%%%%%%%%%%%
\section{Introduction}
%%%%%%%%%%%%%%%%%%%%%%%%%%%%%%%%%%%%%%%%%%%%%%%%%%%%%%%%%%%%%%%%%%%

With the Large Hadron Collider (LHC) successfully running since 2009, the two 
mayor experiments, ATLAS and CMS, discovered a new particle compatible with the
Higgs boson of the Standard Model, with a mass of 126 GeV \cite{:2012gk,:2012gu}.
In addition, they have been collecting data and looking for 
signals beyond the Standard Model (SM). In the case of supersymmetry,
mainly R-Parity conserving models have been considered. Among the last ones, 
in ref.~\cite{Aad:2011cwa} a search for supersymmetric events with two leptons
(electrons and/or muons) was made. These events are produced from decays 
of heavy neutralinos or charginos into the lightest neutralino (lightest 
supersymmetric particle, LSP) via a slepton. In a simplified model where the 
slepton has a mass half way between the heavy neutralino/chargino and the LSP,
a 200 GeV bound on the chargino mass is obtained.
In ref.~\cite{Aad:2011zj} a search for neutralino decaying into a gravitino and 
a photon was performed in an R-Parity conserving gauge mediated supersymmetry 
breaking model. Looking for di-photon events with missing energy, a limit of 805 
GeV for the gluino mass when the neutralino is heavier than 50 GeV was set. In
ref.~\cite{Aad:2011cw} a sbottom pair was searched, with each sbottom decaying 
into a bottom quark and a neutralino. No signal was observed setting a bound 
of 60 GeV for the neutralino mass and 390 GeV for the sbottom mass.
Other R-Parity conserving searches for supersymmetry can be found in
ref.~\cite{Aad:2012zn} for ATLAS and ref.~\cite{Chatrchyan:2011wc} for CMS.
These results mainly point to upper limits for gluino and squark masses which
are starting to reach the 1 TeV level.

In the R-Parity violating scenario, in ref.~\cite{Aad:2011zb} a displaced 
vertex is looked for in association with a muon, found in the decay of a 
neutralino. The non-observation of an excess placed limits on the production 
cross section as a function of the neutralino lifetime. In 
ref.~\cite{Collaboration:2011qr} a massive particle decaying into an electron
and a muon is searched, with no excess found. This places limits on a 
stau that decays via trilinear R-Parity violating couplings, setting a
model dependent lower bound on the stau mass of 1.32 TeV. 
In ref.~\cite{ATLAS:2011ad} supersymmetry was searched in the channel with
jets and an isolated lepton without the observation of an excess. This was 
interpreted in the Bilinear R-Parity Violation (BRpV) model in the tree-level 
dominance scenario, and exclusion limits where set in the gaugino-scalar 
mass plane ($M_{1/2}-m_0$). For example, the gaugino mass has to be larger
than 340 GeV for low scalar masses, and for low gaugino masses the scalar 
mass has a lower bound that reach around 900 GeV.

Although the LHC experiments have not found evidence for supersymmetry, the 
available experimental information is ruling out important sectors of the 
theory, specially where colored particles are light. At least two 
mayor scenarios are not challenged yet: Split Supersymmetry (SS) and R-Parity 
violation.

First, in Split supersymmetry \cite{ArkaniHamed:2004fb,Giudice:2004tc} all
scalar particles are very heavy, with the exception of a SM-like Higgs boson.
The idea behind this model is to keep two of the best phenomenological 
features of supersymmetric theories, namely the unification of gauge 
couplings and the existence of a viable dark matter candidate. The price
to pay is the abandon of the supersymmetric solution to the naturalness 
problem. Since squark are not seen yet at the LHC, this scenario 
acquires strength. Second, although signals for R-Parity have been searched
for, the constraints are strongly model dependent. In BRpV for example,
the search reported in \cite{ATLAS:2011ad} is interpreted in BRpV within
the tree-level dominance scenario. In this case, tree level contributions
to neutrino mass matrix dominates over the one-loop corrections. But this
needs not to be the case, and in other scenarios the interpretation of the
search results would have to be re-done. In fact, in a different scenario 
the decay of the neutralino into a muon and two jets could be very suppressed.

In the original version of Split Supersymmetry, where only one Higgs doublet
remains light, it is not sufficient to add BRpV to generate acceptable 
neutrino masses and mixing angles. One way to fix this problem would be
the inclusion of a gravity motivated term to the neutrino mass matrix
\cite{Berezinsky:2004zb}. A different approach is to keep two Higgs doublets 
light with all the sfermions masses at a high scale, scenario known as Partial
Split Supersymmetry (PSS) \cite{Diaz:2006ee,Diaz:2009gf,Cottin:2011fy}. 
In this scenario, 
despite the fact that the lightest neutralino is unstable due to the presence 
of BRpV, the gravitino is a viable dark matter candidate 
\cite{Roszkowski:2004jd,Diaz:2011pc}.

In PSS, after integrating out the sfermions, new couplings are generated 
between the gauginos, higgsinos, and light Higgs bosons. These couplings 
are called $\tilde g_u$, $\tilde g_d$, $\tilde g'_u$, and $\tilde g'_d$, 
and have boundary conditions at the split supersymmetric scale $\widetilde m$
that relate them to the gauge couplings. They run independently to the 
weak scale acquiring values that differs from coupling to coupling, 
running that has been neglected up to now. In this article we find the 
RGE for these couplings in PSS, and estimate the effect they produce in 
the neutrino observables.

%%%%%%%%%%%%%%%%%%%%%%%%%%%%%%%%%%%%%%%%%%%%%%%%%%%%%%%%%%%%%%%%%%
\section{Electroweak Symmetry Breaking}
%%%%%%%%%%%%%%%%%%%%%%%%%%%%%%%%%%%%%%%%%%%%%%%%%%%%%%%%%%%%%%%%%%

In PSS the Higgs sector is the same as in the MSSM, with two Higgs doublets
$H_u$ and $H_d$, each acquiring vacuum expectation values $v_u$ and $v_d$
respectively. The Higgs potential is also equal to the one in the MSSM.
The presence of BRpV implies that the sneutrinos acquire a vev, 
$v_i$, $i=1,2,3$, one for each generation of sneutrinos. If we do not decouple
yet the sfermions, the minimization conditions for the scalar potential
are the ones in the MSSM-BRpV model \cite{Hirsch:2000ef}. The tadpole condition
for the two Higgs boson vevs are given by,
\begin{eqnarray}
\left( m^2_{H_d}+\mu^2 \right) v_d + v_d D - B_\mu v_u + 
\mu \vec v \cdot \vec\epsilon &=& 0
\nonumber\\
-B_\mu v_d + \left( m^2_{H_u}+\mu^2 \right) v_u - v_u D + 
\vec v\cdot\vec B_\epsilon + v_u \vec\epsilon^{\,2} &=& 0
\label{tadpoleud}
\end{eqnarray}
while the three tadpole conditions associated to the sneutrino vevs are 
condensed into the following equation,
\begin{equation}
v_i D + \epsilon_i \left( -\mu v_d+\vec v\cdot\vec\epsilon \right) +
v_u B_\epsilon^i + v_i M^2_{L_i} = 0
\label{tadpolei}
\end{equation}
Most of the notation in these equations is the usual one, although we note that
$\epsilon_i$ are the supersymmetric BRpV mass terms in the superpotential, 
and that $B_\epsilon^i$ are the bilinear soft terms associated to the former,
analogous to the term $B_\mu$ which is related to the CP-odd Higgs mass
\cite{Diaz:2006ee}.

In PSS sfermions masses are pushed up to a scale $\widetilde m\gg m_Z$, and
eq.~(\ref{tadpolei}) implies that $B_\epsilon^i\sim(v_i/v_u)M^2_{L_i}$,
assuming that the other terms are smaller. This can be achieved in 
theoretical models as noted in ref.~\cite{Diaz:2011pc}. On the other hand, the 
two conditions in eq.~(\ref{tadpoleud}) can be met for a given set of vacuum
expectation values if appropriate values of $m^2_{H_d}$ and $m^2_{H_u}$ are
chosen, solving for these masses in the tadpole conditions. Even if the 
CP-odd Higgs mass is larger than $m_Z$ (but much smaller than $\widetilde m$), 
the first condition in eq.~(\ref{tadpoleud}) is fulfilled if $m_{H_d}\sim m_A$.
In addition, even in the case where the split supersymmetric scale is very large,
eq.~(\ref{tadpoleud}) can be satisfied if $m_{H_u}\sim(v_i/v_u)M_{L_i}$.
All these considerations imply that the scalar potential can have a minimum
with a given set of vevs for the case of very large sfermion masses, provided 
the soft mass parameters satisfy the tadpole equations. Under these conditions,
the mass eigenstate sfermions are integrated out. In particular, the 
sneutrino mass eigenstates that are integrated out have, by construction, a 
higgs component. Conversely, the remaining Higgs bosons fields at the low scale, 
have a sneutrino component. This is crucial for the neutrino mass generation
mechanism to work.

%%%%%%%%%%%%%%%%%%%%%%%%%%%%%%%%%%%%%%%%%%%%%%%%%%%%%%%%%%%%%%%%%%
\section{Neutral fermions in Partial SPLIT SUSY}
%%%%%%%%%%%%%%%%%%%%%%%%%%%%%%%%%%%%%%%%%%%%%%%%%%%%%%%%%%%%%%%%%%

Now we work in the low energy effective model where the sfermions are 
integrated out. As in any supersymmetric model with BRpV, in PSS gauginos and 
higgsinos mix with neutrinos forming a $7\times7$ mass matrix. In the base 
$\psi_0^T=(-i\widetilde B, i\widetilde W^0,\widetilde H_d^0, 
\widetilde H_u^0,\nu_e, \nu_\mu, \nu_\tau)$
we group the mass terms in the lagrangian
\begin{equation}
{\cal L}_N = -\frac{1}{2}\psi_0^T {\cal M}_N^{PSS} \psi_0
\end{equation}
where we divide the mass matrix into four blocks,
\begin{equation}
{\cal M}_N^{PSS}=\left[\begin{array}{cc} {\mathrm M}_{\chi^0}^{PSS} & (m^{PSS})^T \\ 
m^{PSS} & 0 \end{array}\right],
\label{X07x7}
\end{equation}
The upper-left block corresponds to the neutralino sector,
\begin{equation}
{\bf M}_{\chi^0}^{PSS}=\left[\begin{array}{cccc}
M_1 & 0 & -\frac{1}{2}\tilde g'_d v_d & \frac{1}{2}\tilde g'_u v_u \\
0 & M_2 & \frac{1}{2}\tilde g_d v_d & -\frac{1}{2}\tilde g_u v_u \\
-\frac{1}{2}\tilde g'_d v_d & \frac{1}{2}\tilde g_d v_d & 0 & -\mu \\
\frac{1}{2}\tilde g'_u v_u & -\frac{1}{2}\tilde g_u v_u & -\mu & 0
\end{array}\right].
\label{X0massmat}
\end{equation}
The form of this mass matrix is analogous to the case in the MSSM, the 
difference being in the Higgs-gaugino-higgsino couplings $\tilde g$, whose
form in the lagrangian are,
\begin{eqnarray}
{\cal L}_{PSS}^{RpC} \owns
-\textstyle{\frac{1}{\sqrt{2}}} H_u^\dagger
(\tilde g_u \sigma\widetilde W + \tilde g'_u\widetilde B)\widetilde H_u
-\textstyle{\frac{1}{\sqrt{2}}} H_d^\dagger
(\tilde g_d \sigma \widetilde W - \tilde g'_d \widetilde B)\widetilde H_d 
+\mathrm{h.c.}
\label{HHinoGino}
\end{eqnarray}
In the MSSM these couplings are equal to the corresponding gauge couplings, thus
the boundary conditions they satisfy at the scale $\widetilde m$ are
\begin{eqnarray}
&& \tilde g_u(\widetilde m) = \tilde g_d(\widetilde m) = g(\widetilde m)
\nonumber\\
&& \tilde g'_u(\widetilde m) = \tilde g'_d(\widetilde m) = g'(\widetilde m)
\end{eqnarray}
Below the scale $\widetilde m$ these couplings are governed by their own RGE,
which are developedñoped for PSS in the Appendix. The mixing between neutralinos 
and neutrinos is given by the block
\begin{equation}
m^{PSS}=\left[\begin{array}{cccc}
-\frac{1}{2} \tilde g'_d b_1v_u & \frac{1}{2} \tilde g_d b_1v_u 
& 0 &\epsilon_1 \cr
-\frac{1}{2} \tilde g'_d b_2v_u & \frac{1}{2} \tilde g_d b_2v_u &0 
& \epsilon_2 \cr
-\frac{1}{2} \tilde g'_d b_3v_u & \frac{1}{2} \tilde g_d b_3v_u &0 
& \epsilon_3
\end{array}\right].
\end{equation}
The relevant terms in the lagrangian that account for this mixing matrix are,
\begin{equation}
{\cal L}_{PSS}^{RpV} =
-\epsilon_i \widetilde H_u^T \epsilon L_i  
\ -\ 
\textstyle{\frac{1}{\sqrt{2}}} b_i H_u^T\epsilon
(\tilde g_d \sigma\widetilde W-\tilde g'_d\widetilde B)L_i 
\ + \ h.c., 
\label{LSS2HDMRpV}
\end{equation}
where we see the supersymmetric mass parameters $\epsilon_i$, which mix higgsinos 
with neutrinos. The Higgs-gaugino-lepton term proportional to $b_i$ are induced 
in the low energy theory after the sleptons are integrated out, contributing to 
the gaugino-neutrino mixing when the Higgs acquire a vacuum expectation value.

The nature of the $b_i$ terms can be understood as follows. Above the scale 
$\tilde m$ the
Higgs scalars gauge eigenstates mix with the sneutrinos gauge eigenstates, both 
the real parts (CP-even) and the imaginary parts (CP-odd). If we call $s_s^i$ 
and $t_s^i$ the component of the $i$-th real-part-sneutrino inside the CP-even 
Higgs bosons mass eigenstates $h$ and $H$ respectively, it has been proved that 
they satisfy $s_s^i\sim -b_i c_\alpha\sim -c_\alpha v_i/v_u$ and 
$t_s^i\sim -b_i s_\alpha\sim -s_\alpha v_i/v_u$, with analogous relations for 
the imaginary-part-sneutrinos \cite{Diaz:2006ee}. Therefore, the presence
of a non-zero $b_i$ term in eq.~(\ref{LSS2HDMRpV}) indicates that the low
energy fields we call Higgs bosons, in fact have a small sneutrino component.
The sneutrinos are not completely decoupled from the low energy theory below 
$\widetilde m$, but continue living inside our Higgs bosons. In addition, the 
fact that $b_i$ is proportional to the sneutrino vev, implies that if the
$SU(2)$ breaking is switch off, the $b_i$ disappears.

If eq.~(\ref{X07x7}) is block diagonalized, a neutrino effective mass matrix
is generated,
\begin{equation}
{\bf M}_\nu^{eff}=-m^{PSS}\,({\mathrm{M}}_{\chi^0}^{PSS})^{-1}\,(m^{PSS})^T=
\frac{ M_1 \tilde g^2_d + M_2 \tilde g'^2_d }{4\det{M_{\chi^0}^{PSS}}}
\left[\begin{array}{cccc}
\Lambda_1^2        & \Lambda_1\Lambda_2 & \Lambda_1\Lambda_3 \cr
\Lambda_2\Lambda_1 & \Lambda_2^2        & \Lambda_2\Lambda_3 \cr
\Lambda_3\Lambda_1 & \Lambda_3\Lambda_2 & \Lambda_3^2
\end{array}\right],
\label{treenumass}
\end{equation}
with,
\begin{equation}
\det{M_{\chi^0}^{SS}}=-\mu^2 M_1 M_2 + \frac{1}{2} v_u v_d\mu \left( 
M_1 \tilde g_u \tilde g_d + M_2 \tilde g'_u \tilde g'_d \right)
+\frac{1}{16} v_u^2 v_d^2
\left(\tilde g'_u \tilde g_d - \tilde g_u \tilde g'_d \right)^2.
\label{detNeut}
\end{equation}
This matrix, whose matrix elements we denote $M_\nu^{ij}=A^{(0)}\Lambda_i\Lambda_j$
with $A^{(0)}$ being the tree-level contribution in eq.~(\ref{treenumass}), has 
only one non-vanishing eigenvalue. Quantum corrections must be included in order 
to generate a solar as well an atmospheric mass difference. The quantum corrected
neutrino effective mass matrix becomes,
\begin{equation}
M_\nu^{ij}=A\Lambda_i\Lambda_j+C\epsilon_i\epsilon_j
\label{Mnuij}
\end{equation}
The $C$ coefficient is generated at one-loop and in PSS, with 
$m_Z\ll m_A \ll \widetilde m$, the important contributions are from loops with
neutral Higgs bosons, resulting in \cite{Diaz:2009gf}
\begin{equation}
C \approx \frac{m_Z^2\sin^22\beta}
{64\pi^2\mu^2s_\beta^2m_A^2}
\sum_{k=1}^4m_{\chi_k^0}
\left(\tilde g_d N_{k2}-\tilde g'_d N_{k1}\right)^2
\label{CinPSS}
\end{equation}
In addition, the $A$ coefficient receives small one-loop corrections we do not
display, and there is a coefficient $B$ that mixes $\Lambda$ and $\epsilon$,
but can be tunned to zero by choosing an appropriate substraction scale.

Notice that the neutrino mass terms satisfy the two requisites a Majorana
neutrino mass should. First, lepton number is violated, as seen in 
eq.~(\ref{LSS2HDMRpV}) by both terms $\epsilon_i$ and $b_i$. Second, this
neutrino mass vanishes in the limit where $SU(2)$ symmetry is restored, as we see
from $\Lambda_i$ definition that $\Lambda_i\rightarrow0$ as $v\rightarrow0$,
and from eq.~(\ref{CinPSS}) that $C\rightarrow0$ in the same limit.
Notice also that the general theorem shown in ref.~\cite{Hirsch:1997vz} 
is also satisfied
in our model. The theorem states that for a Majorana neutrino mass to be non-zero
in a supersymmetric model, a "Majorana" sneutrino mass should be present also.
This means a mass splitting between the real and imaginary sneutrino masses.
Remembering that our sneutrinos live in the low energy Higgs fields, this
means that the contribution from the Higgs loops should vanish if 
$m_H\rightarrow m_A$. This is satisfied in our model because the above limit
is equivalent to $m_A\gg m_Z$, and from eq.~(\ref{CinPSS}) we see that the
$C$ term goes zero in this limit.

One of the neutrinos remains massless, and the experimental result
$\Delta m^2_{sol}/\Delta m^2_{atm}\approx 0.035$ implies $m_{\nu_3}\gg m_{\nu_2}$
implies,
\begin{eqnarray}
\Delta m^2_{atm} &\approx& \left( A|\vec\Lambda|^2+C|\vec\epsilon\,|^2 \right)^2 -
2AC|\vec\Lambda\times\vec\epsilon|^2,
\nonumber\\
\Delta m^2_{sol} &\approx& \frac{A^2C^2|\vec\Lambda\times\vec\epsilon\,|^4}
{\left( A|\vec\Lambda|^2+C|\vec\epsilon|^2 \right)^2}.
\label{numass2}
\end{eqnarray}
Further approximated results can be obtained if one of the terms $A|\vec\Lambda|^2$
and $C|\vec\epsilon\,|^2$ dominates over the other. The tree-level dominance scenario
is the most commonly assumed \cite{ATLAS:2011ad}, and corresponds to the case 
$A|\vec\Lambda|^2 \gg C|\vec\epsilon\,|^2$.

%%%%%%%%%%%%%%%%%%%%%%%%%%%%%%%%%%%%%%
\section{Numerical Results}
%%%%%%%%%%%%%%%%%%%%%%%%%%%%%%%%%%%%%%

%%%%%%%%%%%%%%%%%%%%%%%%%%%%%%%%%%%%%%
\subsection{Running Couplings in PSS}
%%%%%%%%%%%%%%%%%%%%%%%%%%%%%%%%%%%%%%

One of the guiding principles of Split Supersymmetry is gauge unification.
It is expected that above the scale $M_{GUT}$ particles interactions are governed
by a gauge theory based on a single gauge group. Therefore we impose the
unification of the three gauge couplings at $M_{GUT}$. The starting point is
at the weak scale,
\begin{equation}
g_1^2=\frac{5}{3}\frac{4\pi\alpha_e}{c_W^2}
\,,\qquad
g_2^2=\frac{4\pi\alpha_e}{s_W^2}
\end{equation}
with $\alpha_e^{-1}(m_Z)=128.962\pm0.014$ \cite{Hoecker:2010qn} and 
$s_W^2(m_Z)=0.23116\pm0.00013$ \cite{Nakamura:2010zzi}.
We run the couplings
$g_1$ and $g_2$ until they meet at a scale we define as $M_{GUT}$. At that point 
we impose $g_3=g_2=g_1$ and run back $g_3$ to the weak scale, where we impose that 
the strong coupling constant satisfy the experimental constraint
$\alpha_s(m_Z)=0.1184\pm0.0007$ \cite{Nakamura:2010zzi}.

\begin{figure}[ht]
\centerline{\protect\vbox{\epsfig{file=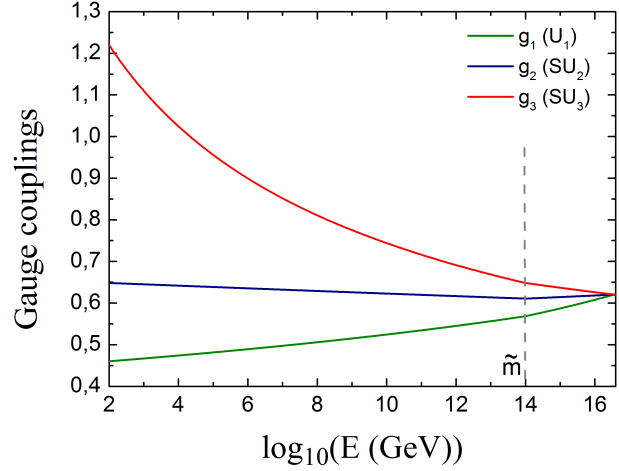,width=0.6\textwidth}}}
\caption{\it Unification of gauge couplings in PSS.
}
\label{gauge}
\end{figure}

As a working example we take $\widetilde m=10^{14}$ GeV and $\tan\beta=10$,
and find $M_{GUT}=3.6\times 10^{16}$ GeV with $\alpha_s(m_Z)=0.1189$, in agreement
with the experimental data. The running of the three gauge coupling constans can be 
seen in Fig.~\ref{gauge}.
%
%%%%%%%%%% FIGURE %%%%%%%%%%%%%%%%%%
%%%%%%%%%%%%%%%%%%%%%%%%%%%%%%%%%%%%
%
The RGE for $g_i$ above the scale $\widetilde m$ are the ones for the MSSM, while
below $\widetilde m$ the RGE to be used are the ones for PSS and they are given in 
the Appendix.
%
%%%%%%%%%% FIGURE %%%%%%%%%%%%%%%%%%
\begin{figure}[ht]
\centerline{\protect\vbox{\epsfig{file=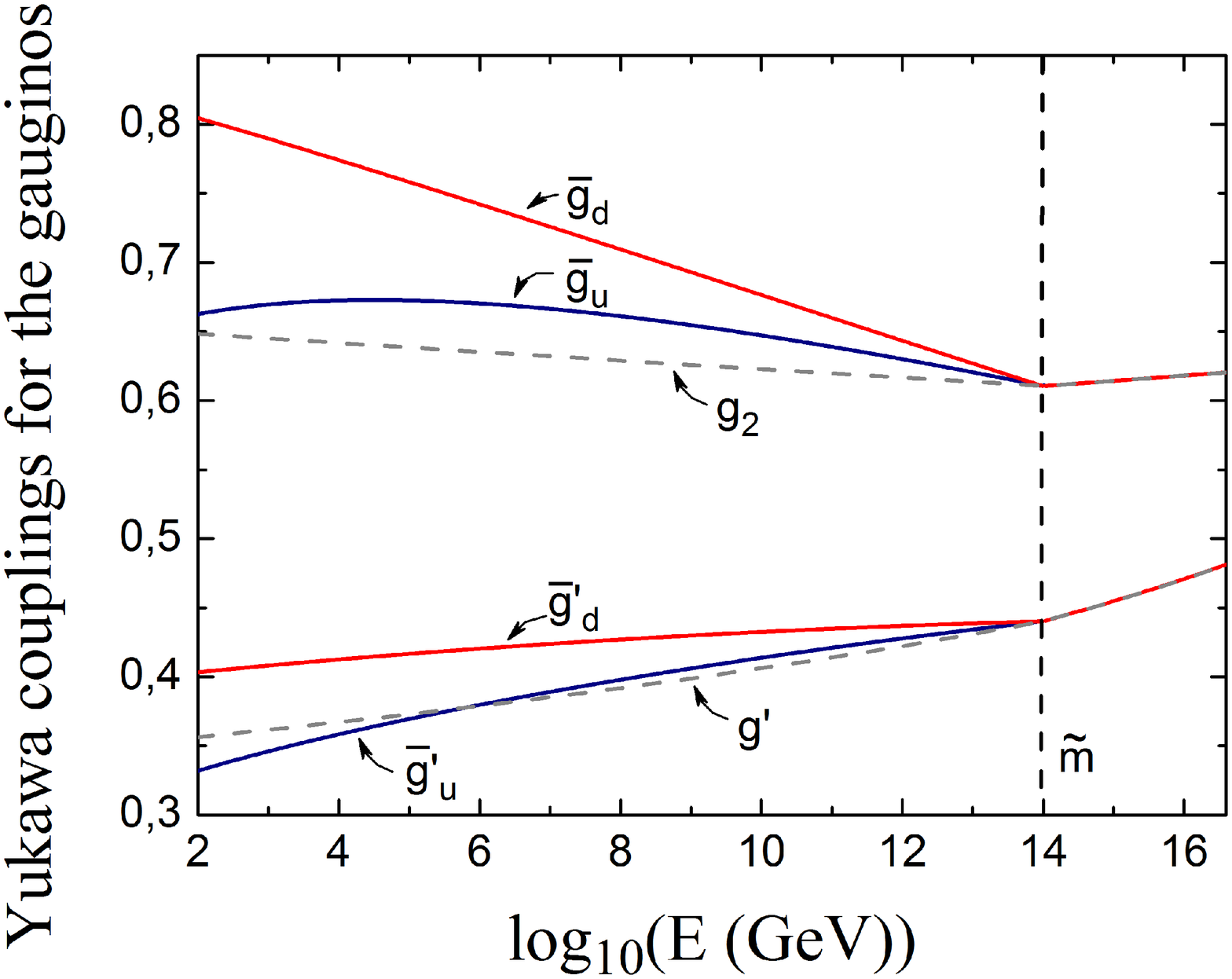,width=0.6\textwidth}}}
\caption{\it Running of Higgs-higgsino-gaugino couplings.
}
\label{gaugino}
\end{figure}
%%%%%%%%%%%%%%%%%%%%%%%%%%%%%%%%%%%%
%
In the same working scenario we plot in Fig.~\ref{gaugino} the Higgs-higgsino-gaugino
couplings $\tilde g$. The top three curves correspond to $\tilde g_d$, $\tilde g_u$, 
and $g$ ($=g_2$), the third one shown for comparison. Above $\widetilde m$ the three
coincide as it should be in the MSSM, and below $\widetilde m$ they separate for
as much as $20\%$ in the case of $\tilde g_d$. The lower three curves correspond to
$\tilde g'_d$, $\tilde g'_u$, and $g'$ ($=g_1\sqrt{3/5}$), and they can differ for
as much as $10\%$ in this scenario. Considering the high precission for the 
measurements of neutrino mass differences, these RGE effects will have an impact
as we will show next.

%%%%%%%%%%%%%%%%%%%%%%%%%%%%%%%%%%%%%%%%%%%%%
\subsection{Neutrino Mass Differences in PSS}
%%%%%%%%%%%%%%%%%%%%%%%%%%%%%%%%%%%%%%%%%%%%%

It is clear that the larger $\widetilde m$ the larger the RGE effects on the 
gaugino couplings $\tilde g$. This is obvious from the definition, but it is also
apparent in Fig.~\ref{gaugino}. This affects the neutrino mass matrix in
eq.~(\ref{Mnuij}) through the coefficients $A$ and $C$, which depend on the gaugino
couplings as can be seen in eqs.~(\ref{treenumass}) and (\ref{CinPSS}). The size and 
shape of the effect depends also on the particular point in parameter space we are
working with. 

First of all we look for a working scenario with neutrino masses and mixing angles 
in agreement with experimental results, when neglecting the RGE effect. This means we
work in the approximation $\tilde g_d=\tilde g_u=g$ and $\tilde g'_d=\tilde g'_u=g'$
at the weak scale.
For the PSS parameters we choose the values indicated in Table  \ref{PSSscenario}.
\begin{table}[ht]
\caption{PSS parameters for the working scenario $S$} % title of Table
\centering % used for centering table
\begin{tabular}{c c c } % centered columns (4 columns)
\hline\hline %inserts double horizontal lines
PSS parameter & $S$ & Units  \\ [0.5ex] % inserts table
%heading
\hline % inserts single horizontal line
 $\tan\beta$ & $10$   & $...$   \\
 $\mu$   & $450$   & GeV  \\
 $M_{2}$ & $300$   & GeV  \\
 $M_{1}$ & $150$   & GeV  \\
 $m_{h}$ & $120$   & GeV  \\
 $m_{A}$ & $1000$  & GeV  \\
 $Q$     & $527$   & GeV  \\
[1ex] % [1ex] adds vertical space
\hline %inserts single line
\end{tabular}
\label{PSSscenario} % is used to refer this table in the text
\end{table}

For the RpV parameters we perform a scan over parameter space according to the 
intervals indicated in the third column in Table \ref{RPVscenario}. 
\begin{table}[ht]
\caption{RpV parameters for the working scenario $S$} % title of Table
\centering % used for centering table
\begin{tabular}{c c c c } % centered columns (4 columns)
\hline\hline %inserts double horizontal lines
RpV parameter & $S$ & Scanned range & Units  \\ [0.5ex] % inserts table
%heading
\hline % inserts single horizontal line
 $\epsilon_{1}$ & $0.0346$ & $ [-1,1]$  & GeV      \\
 $\epsilon_{2}$ & $0.2516$ & $ [-1,1]$  & GeV      \\
 $\epsilon_{3}$ & $0.3504$ & $ [-1,1]$  & GeV      \\
 $\Lambda_{1}$ & $0.0348$  & $ [-1,1]$  & $\mbox{GeV}^2$     \\
 $\Lambda_{2}$ & $-0.0021$ & $ [-1,1]$  & $\mbox{GeV}^2$     \\
 $\Lambda_{3}$ & $0.0709$  & $ [-1,1]$  & $\mbox{GeV}^2$     \\
[1ex] % [1ex] adds vertical space
\hline %inserts single line
\end{tabular}
\label{RPVscenario} % is used to refer this table in the text
\end{table}
For each scanned point we calculate (See ref.~\cite{Maltoni:2004ei} for the 
status of best fits to neutrino parameters)
\begin{equation}
\chi^2 = \left(\frac{10^3 \Delta m_{atm}^{2} -2.4}{0.4}    \right)^2 + \left(\frac{10^5 \Delta m_{sol}^{2}- 7.7}{0.6}    \right)^2 + \left(\frac{\sin^2 \vartheta_{atm}-0.505}{0.165}    \right)^2 + \left(\frac{\sin^2 \vartheta_{sol}-0.33}{0.07}    \right)^2  
\end{equation}
and demand $\chi^2<1$. Among all the solutions we choose the one given by the 
second column in Table \ref{RPVscenario} with $\chi^2=0.072$, and refer to it as 
scenario $S$. This is an example of what we call a one-loop dominated solution, 
since $|A\vec\Lambda^2 / (C\vec\epsilon\,^2)| = 0.29$. 
Without approximations in the neutrino mass matrix in eq.~(\ref{Mnuij}) we obtain 
the following observables,
\begin{eqnarray}
\Delta m_{\mbox{atm}}^2 &=& 2.4 \times 10^{-3 } \; \mbox{eV}^2
\nonumber\\
\Delta m_{\mbox{sol}}^2 &=& 7.73 \times 10^{-5} \; \mbox{eV}^2
\nonumber\\
\sin ^2 \vartheta_{\mbox{atm}} &=& 0.4577
\nonumber\\
\sin ^2 \vartheta_{\mbox{sol}} &=& 0.3337
\label{SPoint}\\
\sin^2 \vartheta_{\mbox{reac}} &=& 2.30 \times 10^{-5}
\nonumber\\
m_{\beta \beta} &=& 0.0029 \; \mbox{eV}
\nonumber
\end{eqnarray}
which are well within experimental constraints. The last parameter corresponds to the
effective Majorana neutrino mass, which must satisfy $m_{\beta \beta}<0.7$ eV 
according to neutrinoless double beta decay experiments. It is useful to confront
these results with approximated neutrino mass differences and angles in the 
one-loop dominated scenario. In this situation we have,
\begin{eqnarray}
\Delta m^2 _{\mbox{sol}} &\approx&   
A^2\frac{|\Lambda \times \epsilon|^4}{|\epsilon|^4}
\nonumber\\
\Delta m^2 _{\mbox{atm}} &\approx&  C^2|\epsilon|^4
\nonumber\\
\tan^2 \vartheta _{\mbox{sol}} &\approx&  \frac{\Lambda_{1}^2 (\epsilon_{2}^2 + \epsilon_{3}^2)}{(\Lambda_{2}\epsilon_{3}- \Lambda_{3}\epsilon_{2})^2}
\\
\tan^2 \vartheta _{\mbox{atm}} &\approx&  \frac{\epsilon_{2}^2}{\epsilon_{3}^2}
\nonumber\\
\tan^2 \vartheta _{\mbox{reac}} &\approx& \frac{\epsilon_{1}^2}{\epsilon_{2}^2 + \epsilon_{3}^2}
\nonumber
\end{eqnarray}
Clearly the mixing angles in first approximation are not affected by the RGE effects,
while the mass differences are. This is confirmed in the following plots.

%
%%%%%%%%%% FIGURE %%%%%%%%%%%%%%%%%%
\begin{figure}[ht]
\centerline{\protect\vbox{\epsfig{file=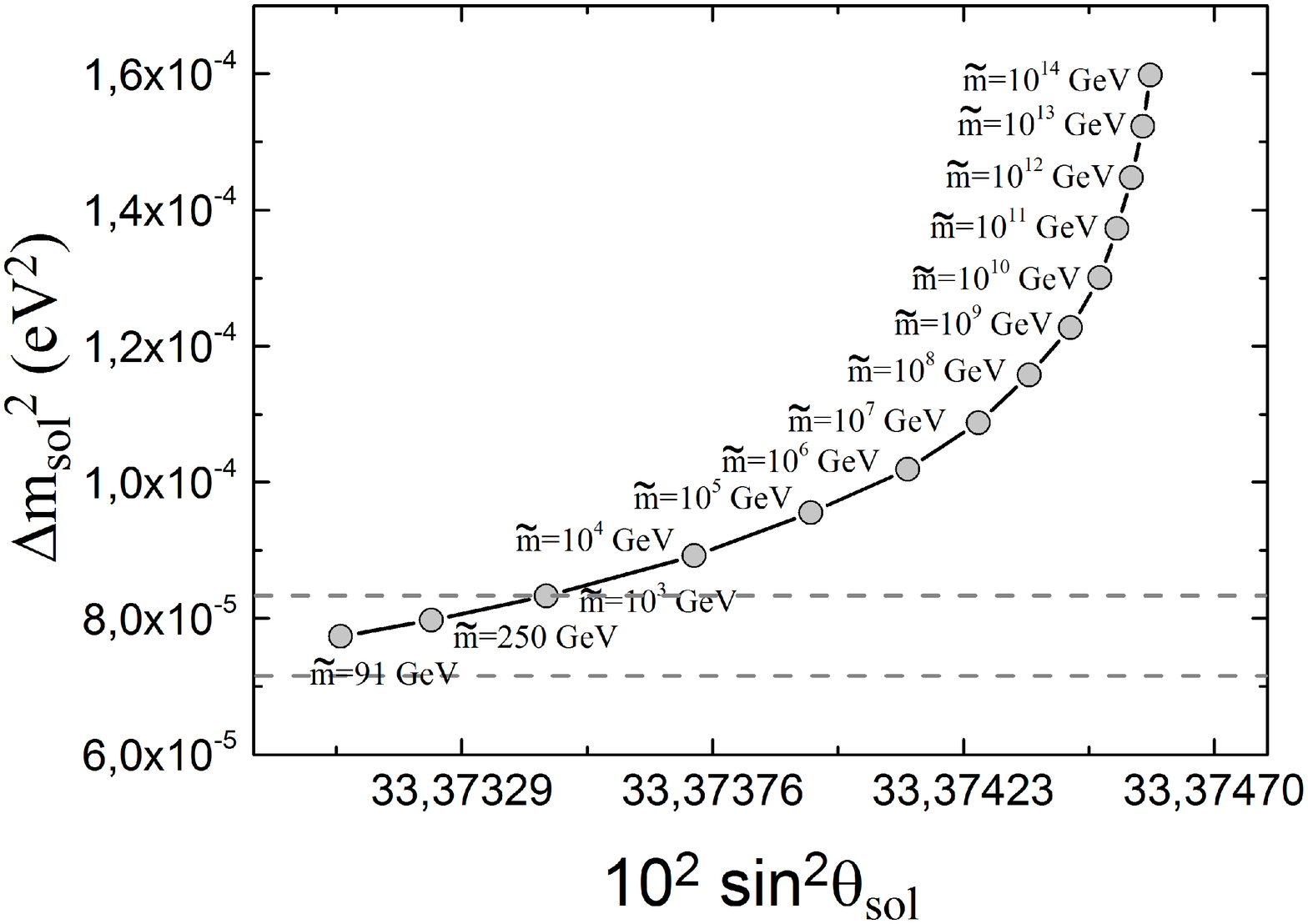,width=0.7\textwidth}}}
\caption{\it Running effects on solar mass and solar angle.
}
\label{matm}
\end{figure}
%%%%%%%%%%%%%%%%%%%%%%%%%%%%%%%%%%%%
%
In Fig.~\ref{matm} we have the atmospheric mass squared difference in the y-axis and
the sine squared of the atmospheric angle in the x-axis. The lowest point corresponds
to $\widetilde m=m_Z$, which means we have the MSSM all the way to the weak scale. 
This is also equivalent to neglect the RGE effects. The $\Delta m^2_{atm}$ and
$\sin^2\theta_{atm}$ values for this case are the ones given in eq.~(\ref{SPoint}).
In the following points we study the effect of the $\tilde g$ running, and each
of them are defined by an increasing value of $\widetilde m$. Two things are immediately
apparent: (i) the effect on the atmospheric angle is negligible, and (ii) the effect 
on the atmospheric mass is very important. Indeed, already at $\widetilde m=10^4$ GeV
the value of $\Delta m^2_{atm}$ leaves the $3\sigma$ allowed interval.

%
%%%%%%%%%% FIGURE %%%%%%%%%%%%%%%%%%
\begin{figure}[ht]
\centerline{\protect\vbox{\epsfig{file=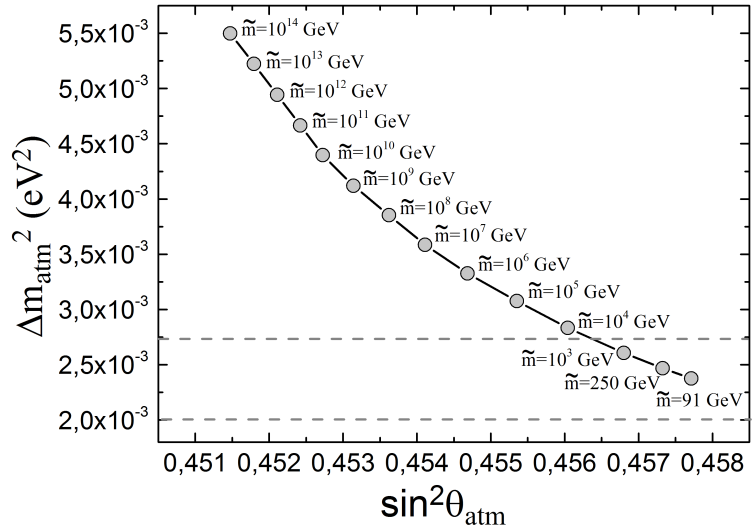,width=0.7\textwidth}}}
\caption{\it Running effects on atmospheric mass.
}
\label{msol}
\end{figure}
%%%%%%%%%%%%%%%%%%%%%%%%%%%%%%%%%%%%
%
In Fig.~\ref{msol} we have a similar plot, with the solar mass squared difference 
in the y-axis and the sine squared of the solar angle in the x-axis. The situation
is similar, with a very small effect on the solar angle and a very large effect
on the solar mass. Indeed, already for $\widetilde m=10^3$ GeV the solar mass
calculated including the RGE effect leaves the $3\sigma$ allowed interval.

%%%%%%%%%%%%%%%%%%%%%%%%%%%%%%%%%%%%%%%%%%%%%%%%%%%%%%%%%%%%%%%%%%%%%%%%%%%%%%%
\section{Summary}
%%%%%%%%%%%%%%%%%%%%%%%%%%%%%%%%%%%%%%%%%%%%%%%%%%%%%%%%%%%%%%%%%%%%%%%%%%%%%%%
\label{conclusions}

We have seen that neutrino masses can be generated in PSS via a low energy
see-saw mechanism, where neutrinos mix with neutralinos through BRpV couplings.
The neutrino masses and mixing angles generated depend, through both the 
tree-level and the one-loop contributions, on the Higgs-gaugino-higgsino 
couplings. These couplings have boundary conditions at the split supersymmetric 
scale $\widetilde m$ that relate them to the gauge couplings. But the RGEs that
govern them are different from the RGEs for the gauge couplings, which implies 
they have different values at the weak scale. We have found the RGE for the 
relevant couplings in PSS, and showed that the effect of their running is 
large enough to affect the neutrino observables. Although mixing angles are not 
affected, we have shown that the 
atmospheric and solar mass squared differences are very sensitive to this 
running, such that with a moderate split supersymmetric scale, it already can 
change the viability of a given scenario.

%%%%%%%%%%%%%%%%%%%%%%%%%%%%%%%%%%%%%%%%%%%%%%%%%%%%%%%%%%%%%%%%%

%%%%%%%%%%%%%%%%%%%%%%%%%%%%%%%%%%%%%%%%%%%%%%%%%%%%%%%%%%%%%%%%%%
\begin{acknowledgments}

{\small 
One of us (F.C.) is grateful to the physics department of the Pontificia Universidad Catolica de Chile for supporting his work. 
This work was partly funded by Conicyt grant 1100837 (Fondecyt Regular).}

\end{acknowledgments}
%%%%%%%%%%%%%%%%%%%%%%%%%%%%%%%%%%%%%%%%%%%%%%%%%%%%%%%%%%%%%%%%%

\newpage

%%%%%%%%%%%%%%%%%%%%%%%%%%%%%%%%%%%%%%%%%%%%%%%%%%%%%%%%%%%%%%%%%%
\section{Appendix: Renormalization Group Equations}
%%%%%%%%%%%%%%%%%%%%%%%%%%%%%%%%%%%%%%%%%%%%%%%%%%%%%%%%%%%%%%%%%%

Here we study the Renormalization Group Equations for gauge, Yukawa, and 
gaugino-Higgs-higgsino couplings in Partial Split Supersymmetry. As usual, we
define $t=\ln Q^2$ with $Q$ the arbitrary scale introduced by dimensional
regularization. We use the results, and follow as close as possible the 
notation, given by \cite{Machacek:1983tz}.

%%%%%%%%%%%%%%%%%%%%%%%%%%%%%%%%%%%%%%%%%%%%%%%%%%%%%%%%%%%%%%%%%%
\subsection{General Two-loop RGE for Gauge Couplings}
%%%%%%%%%%%%%%%%%%%%%%%%%%%%%%%%%%%%%%%%%%%%%%%%%%%%%%%%%%%%%%%%%%

In a $SU(3)\times SU(2)\times U(1)$ gauge theory, the three gauge couplings 
have the following RGE,
\begin{equation}
\beta_{g_i}=\frac{dg_i}{dt} = \frac{1}{16\pi^2} g_i^3 b_i
+\frac{1}{(16\pi^2)^2} g_i^3 \Big\{ B_{ij} g_j^2 - 2 Y_i(F) \Big\}
\label{RGEgi}
\end{equation}
where there is a sum over $j$ but not over $i$.
Here $i=1,2,3$ refers to the $U(1)$, $SU(2)$, and $SU(3)$ groups respectively,
denoted in general as $G_i$, with
$g_i$ being the corresponding gauge coupling. The term proportional to $b_i$ is
the one-loop contribution, with
\begin{equation}
b_i = -\frac{11}{3}C_2(G_i)+\frac{2}{3}\sum_f T(R_i^f)d(R_j^f)d(R_k^f)+
\frac{1}{3}\sum_s T(R_i^s)d(R_j^s)d(R_k^s)
\label{RGEga}
\end{equation}
and $i\ne j\ne k$. The first term is the contribution from the gauge bosons, with 
$C_2(G_i)$ the normalization factor of the quadratic Casimir operator $C_2$ for the 
adjoint representation of the group $G_i$. This Casimir normalization is defined as,
\begin{eqnarray}
C_2 &=& A_i^a A_i^a = C_2(G_i) {\bf 1}
\label{casimir}
\end{eqnarray}
also satisfying,
\begin{eqnarray}
{\mathrm{Tr}}(A_i^a A_i^b) &=& C_2(G_i) \delta^{ab}
\label{TrAA}
\end{eqnarray}
where $A_i^a$ are the $N$ generators in the adjoint representation of the group 
$G_i$, labeled by the indices $a,b=1,...N$. For $G_N=SU(N)$, $N>1$ we have 
$C_2(G_N)=N$, while for $G_1=U(1)$ we have $C_2(G_1)=0$. In particular, for the 
case of $SU(2)$ and $SU(3)$ we have the two well known results,
\begin{eqnarray}
\epsilon_{acd}\epsilon_{bcd}=2\delta_{ab}\,,\quad &{\mathrm{with}}& \qquad
(A_2^a)_{bc}=\epsilon_{abc}
\nonumber\\
f_{acd}f_{bcd}=3\delta_{ab}\,,\quad &{\mathrm{with}}& \qquad
(A_3^a)_{bc}=f_{abc}
\end{eqnarray}
with $\epsilon_{abc}$ and $f_{abc}$ the structure constants of the $SU(2)$ and
$SU(3)$ Lie groups respectively.

The second term in eq.~(\ref{RGEga}) corresponds to the contribution from fermions 
living in a general representation $R_i^f$, where $f$ runs over all different 
fermions and $i=1,2,3$ refers to the gauge group $G_i$ as before. In our model, 
fermions live in the fundamental representation, whose generators we call 
$F_i^a$. They are equal to half the Pauli matrices in the case of $SU(2)$,
$F_2^a=\sigma^a/2$, $a=1,...3$, and half the Gelmann matrices in the case of 
$SU(3)$, $F_3^a=\lambda^a/2$, $a=1,...8$. They satisfy,
\begin{equation}
{\mathrm{Tr}}(F_i^a F_i^b) = T(R_i^f) \delta^{ab} = \frac{1}{2} \delta^{ab}
\end{equation}
where $T(R_i^f)$ is known as the Dynkin index of the representation $R_i^f$.
In the case of the fundamental representation the conventional normalization is
$T(R_i^f)=1/2$. This last equation is analogous to eq.~(\ref{TrAA}). Finally,
factors $d(R_j^f)$ and $d(R_k^f)$ are the dimensions of the fermion multiplet in
the other two groups $j,k\ne i$. In the case of $U(1)$ the Dynkin index of the 
fundamental representation is $T(R_1^f)=y^2$, with $y$ the appropriately 
normalized hypercharge.

The third term in eq.~(\ref{RGEga}) is the contribution from scalars when they 
live in the general representation $R_i^s$, where $s$ runs over all the scalars. 
If this representation is the fundamental, of course we have $T(R_i^s)=1/2$.

We turn now to the two-loop contributions to the RGE for the gauge couplings,
given in the second term of eq.~(\ref{RGEgi}). The term $B_{ij}$ inside the 
bracket is equal to,
\begin{eqnarray}
B_{ii} &=& -\frac{34}{3}\left[C_2(G_i)\right]^2 +
\sum_f \left[ \frac{10}{3} C_2(G_i) + 2C_2(R_i^f) \right]
T(R_i^f)d(R_j^f)d(R_k^f)
\nonumber\\
&& \qquad\qquad\qquad\,\, + \sum_s \left[ \frac{2}{3} C_2(G_i) + 4C_2(R_i^s) \right]
T(R_i^s)d(R_j^s)d(R_k^s)
\nonumber\\
B_{ij} &=& 
\sum_f 2 T(R_i^f) C_2(R_j^f) d(R_j^f)d(R_k^f) +
\sum_s 4 T(R_i^s) C_2(R_j^s) d(R_j^s)d(R_k^s)
\label{Bijall}
\end{eqnarray}
with $i\ne j\ne k$ and there is no sum over repeated $i$ indices. Here we see
the quadratic Casimir normalization factor evaluated in the fundamental
representation. The version of eq.~(\ref{casimir}) for the fundamental
representation is,
\begin{eqnarray}
C_2 &=& F_i^a F_i^a = C_2(R_i) {\bf 1}
\label{casimirFund}
\end{eqnarray}
with an upper index $f$ or $s$ on $R_i$ for the fermion and scalar cases. 
It can be easily shown the following relation with the Dynkin index,
\begin{equation}
C_2(R_i) d(R_i) = T(R_i) d(G_i)
\end{equation}
This implies for the fundamental representation of $SU(N)$ that 
$C_2(R_N)=(N^2-1)/(2N)$. For the case of $U(1)$ we have $C_2(R_1)=T(R_1)=y^2$,
with $y$ defined before as the normalized hypercharge.

The $Y_i(F)$ term inside the bracket in eq.~(\ref{RGEgi}) is given by,
\begin{equation}
Y_i(F) = \frac{1}{d(G_i)} \sum_f {\mathrm{tr}} \left[
C_2(R_i^f) N_c Y_f^a Y_f^{a\dagger} \right]
\label{YiF}
\end{equation}
where $Y_f^a$ are Yukawa-type terms (see Appendix C), $f$ are all the 
fermions that couple through that coupling, and $a$ labels the real scalars 
coupled to those fermions.

%%%%%%%%%%%%%%%%%%%%%%%%%%%%%%%%%%%%%%%%%%%%%%%%%%%%%%%%%%%%%%%%%%
\subsection{Explicit Two-loop RGE for Gauge Couplings in PSS}
%%%%%%%%%%%%%%%%%%%%%%%%%%%%%%%%%%%%%%%%%%%%%%%%%%%%%%%%%%%%%%%%%%

Here we explicitly apply the above formulae to our model. We find first 
the one loop parameters $b_i$, which for $SU(3)$ in PSS is,
\begin{equation}
b_3 = \bigg\{-\frac{11}{3}\times 3
+\frac{2}{3}\times\frac{1}{2}\times n_g\left(2+1+1\right) \bigg\}_{SM}
+\frac{2}{3}\times3 = -5
\end{equation}
which includes contributions from gluons, $Q$, $u_R$, $d_R$, and $\tilde g$. Note 
that $n_g$ is the number of generations, that the gluinos live in the adjoint 
representation thus we use $T(R^f_3)=3$, and that PSS does not include squarks.
For convenience we include in brackets the SM contribution to the RGE.

Second, for $SU(2)$ we have,
\begin{equation}
b_2=\bigg\{ -\frac{11}{3}\times 2
+\frac{2}{3}\times\frac{1}{2}\times n_g\left(3+1\right)
+ \frac{1}{3} \times \frac{1}{2} \bigg\}_{SM}
+\frac{2}{3}\times\Big[2+\frac{1}{2}(1+1)\Big]
+ \frac{1}{3} \times \frac{1}{2} = -1
\end{equation}
which includes the contributions from $W$, $Q$, $L$, $H_d$, $\widetilde W$, 
$\widetilde H_u$, $\widetilde H_d$, and $H_u$. Note that the winos live 
in the adjoint representation, thus we take $T(R_2^f)=2$.

Third, we obtain for $U(1)$,
\begin{eqnarray}
b_1 &=& \Bigg\{ -\frac{11}{3}\times 0
+\frac{2}{3}\times\frac{3}{20}\times n_g\Big[(1/3)^2\times 2\times 3
+(2/3)^2\times 3+(-4/3)^2\times 3 +
\nonumber\\ && \qquad\qquad\qquad\qquad\qquad\qquad
(-1)^2\times 2+(2)^2 \Big]
+\frac{1}{3}\times\frac{3}{20}\times(-1)^2\times 2
\Bigg\}_{SM}
\nonumber\\ &&
+\frac{2}{3}\times\frac{3}{20}\times\Big[ (1)^2\times 2+(-1)^2\times 2 \Big]
+\frac{1}{3}\times\frac{3}{20}\times(1)^2\times 2 = \frac{23}{5}
\end{eqnarray}
which includes contributions from $B$, $Q$, $d_R$, $u_R$, $L$, $e_R$, $H_d$, 
$\widetilde H_u$, $\widetilde H_d$, and $H_u$. Note that the quadratic Casimir 
for the adjoint
representation in $U(1)$ is null, and that the factor $3/20$ is the hypercharge 
normalization. We summarize the values of $b_i$ in Table \ref{bi} for both the
SM and PSS. The corresponding values for Split Supersymmetry can be found in
ref.~\cite{Giudice:2004tc}.
%
%------------------------ TABLE  -------------------------------------
\begin{table}[ht]
\begin{center}
\caption{$b_i$ parameters in the one-loop RGE for gauge couplings.}
\bigskip
\begin{minipage}[t]{0.8\textwidth}
\begin{tabular}{cccc}
\hline
Model           & $b_1$           & $b_2$           & $b_3$     \\
\hline \hline
SM              & $\frac{41}{10}$ & $-\frac{19}{6}$ & $-7$      \\
\hline
PSS             & $\frac{23}{5}$  & $-1$            & $-5$      \\
\hline \hline \label{bi}
\end{tabular}
\end{minipage}
\end{center}
\end{table}
%---------------------------------------------------------------------

Next we calculate the two-loop parameters $B_{ij}$ in eq.~(\ref{RGEgi}).
The diagonal terms $B_{ii}$ are calculated from eq.~(\ref{Bijall}),
\begin{eqnarray}
B_{11} &=& \Bigg\{-\frac{34}{3}\times0 +
2n_g \Big(\frac{3}{20}\Big)^2 \bigg[(1/3)^4 \times 3\times 2 +
(2/3)^4 \times 3 + (-4/3)^4 \times 3 + (-1)^4 \times 2 + (2)^4 \bigg] 
\nonumber\\ && \quad
+ 4 \Big(\frac{3}{20}\Big)^2 (-1)^4 \times 2
\Bigg\}_{SM} + 2 \Big(\frac{3}{20}\Big)^2 \Big[ (1)^4 \times 2 + 
(-1)^4 \times 2 \Big] 
+ 4 \Big(\frac{3}{20}\Big)^2 (-1)^4 \times 2
\nonumber\\ &=& \frac{217}{50}
\end{eqnarray}
where we are including $B$, $Q$, $d_R$, $u_R$, $L$, $e_R$, $H_d$,
$\widehat H_u$, $\widehat H_d$, and $H_u$;
\begin{eqnarray}
B_{22} &=& \Bigg\{-\frac{34}{3}(2)^2 + n_g 
\bigg[ \frac{10}{3}\times 2 +2\times \frac{3}{4} \bigg]
\times \bigg( \frac{1}{2} \times 3 + \frac{1}{2} \bigg) +
\bigg[ \frac{2}{3}\times 2 +4\times \frac{3}{4} \bigg] \times \frac{1}{2}
\Bigg\}_{SM}
\\ &+&
\bigg[ \frac{10}{3}\times 2 +2\times \frac{3}{4} \bigg] \times 
\bigg( \frac{1}{2} + \frac{1}{2} \bigg) +
\bigg[ \frac{10}{3}\times 2 +2\times 2 \bigg] \times 2 +
\bigg[ \frac{2}{3}\times 2 +4\times \frac{3}{4} \bigg] \times \frac{1}{2}
= \frac{225}{6}
\nonumber
\end{eqnarray}
where we have included $W$, $Q$, $L$, $H_d$, $\widetilde H_u$, 
$\widetilde H_d$, $\widetilde W$, and $H_u$; and
\begin{eqnarray}
B_{33} &=& \Bigg\{-\frac{34}{3}(3)^2 + n_g 
\bigg[ \frac{10}{3}\times 3 +2\times \frac{4}{3} \bigg]
\times \bigg[ \frac{1}{2} \times 2 + \frac{1}{2} + \frac{1}{2}
\bigg] \Bigg\}_{SM} 
\nonumber\\ && \qquad\qquad\qquad\,\,\,
+ \bigg[ \frac{10}{3}\times 3 +2\times 3 \bigg] \times 3 = 22
\end{eqnarray}
where we have included $g$, $Q$, $u_R$, $d_R$, and $\tilde g$.

The off-diagonal terms $B_{ij}$ are also calculated from eq.~(\ref{Bijall}),
\begin{eqnarray}
B_{12} &=& \Bigg\{ 2n_g \times \frac{3}{4} \times \frac{3}{20} 
\bigg[ (1/3)^2\times 2\times 3 + (-1)^2\times 2 \bigg] +
4\times \frac{3}{4} \times \frac{3}{20} (-1)^2 \times 2
\Bigg\}_{SM} 
\nonumber\\ &&
+2 \times\frac{3}{4} \times \frac{3}{20} \bigg[ (1)^2\times 2 + (-1)^2\times 2 \bigg]
+4 \times \frac{3}{4} \times \frac{3}{20} (1)^2 \times 2
= \frac{9}{2}
\end{eqnarray}
\begin{eqnarray}
B_{21} &=& \Bigg\{ 2n_g \times \frac{1}{2} \times \frac{3}{20} 
\bigg[ (1/3)^2 \times 3 + (-1)^2 \bigg] +
4\times \frac{1}{2} \times \frac{3}{20} (-1)^2
\Bigg\}_{SM} 
\nonumber\\ &&
+2 \times\frac{1}{2} \times \frac{3}{20} \bigg[ (1)^2 + (-1)^2 \bigg]
+4 \times \frac{1}{2} \times \frac{3}{20} (1)^2
= \frac{3}{2}
\end{eqnarray}
where in both $B_{12}$ and $B_{21}$ we included $Q$, $L$, $H_d$, $\widetilde H_u$, 
$\widetilde H_d$, and $H_u$;
\begin{eqnarray}
B_{13} &=& \Bigg\{ 2n_g \times \frac{4}{3} \times \frac{3}{20} 
\bigg[ (1/3)^2\times 2\times 3 + (2/3)^2\times 3 + (-4/3)^2\times 3 \bigg]
\Bigg\}_{SM} = \frac{44}{5}
\end{eqnarray}
\begin{eqnarray}
B_{31} &=& \Bigg\{ 2n_g \times \frac{1}{2} \times \frac{3}{20} 
\bigg[ (1/3)^2\times 2 + (2/3)^2 + (-4/3)^2 \bigg]
\Bigg\}_{SM} = \frac{11}{10}
\end{eqnarray}
where in both $B_{13}$ and $B_{31}$ we included $Q$, $d_R$, and $u_R$; and
\begin{eqnarray}
B_{23} &=& \Bigg\{ 2n_g \times\frac{4}{3} \times\frac{1}{2} \times 3
\Bigg\}_{SM} = 12
\end{eqnarray}
\begin{eqnarray}
B_{32} &=& \Bigg\{ 2n_g \times\frac{3}{4} \times\frac{1}{2} \times 2
\Bigg\}_{SM} = \frac{9}{2}
\end{eqnarray}
where in both $B_{23}$ and $B_{32}$ we included only $Q$.
We summarize the two-loop parameters $B_{ij}$ in Table \ref{Bij}.
%
%------------------------ TABLE  -------------------------------------
\begin{table}[ht]
\begin{center}
\caption{$B_{ij}$ parameters in the two-loop RGE for gauge couplings.}
\bigskip
\begin{minipage}[t]{0.8\textwidth}
\begin{tabular}{cccccccccc}
\hline
Model           & $B_{11}$         & $B_{12}$        & $B_{13}$       & $B_{21}$       &
$B_{22}$        & $B_{23}$         & $B_{31}$        & $B_{32}$       & $B_{33}$       \\
\hline \hline
SM              & $\frac{199}{50}$ & $\frac{27}{10}$ & $\frac{44}{5}$ & $\frac{9}{10}$ &
$\frac{35}{6}$  & $12$             & $\frac{11}{10}$ & $\frac{9}{2}$  & $-26$          \\
\hline
PSS             & $\frac{217}{50}$ & $\frac{9}{2}$   & $\frac{44}{5}$ & $\frac{3}{2}$  &
$\frac{225}{6}$ & $12$             & $\frac{11}{10}$ & $\frac{9}{2}$  & $22$           \\
\hline \hline \label{Bij}
\end{tabular}
\end{minipage}
\end{center}
\end{table}
%---------------------------------------------------------------------

Finally we calculate the two-loop terms $Y_i(F)$ given in eq.~(\ref{YiF}). For
the $U(1)$ gauge group we have,
\begin{eqnarray}
Y_1(F) &=& \Bigg\{ 3 \times \frac{3}{20} \bigg[
(1/3)^2 \, {\mathrm{Tr}}\left( Y_uY_u^\dagger + Y_dY_d^\dagger \right)
+ (2/3)^2 \, {\mathrm{Tr}}\left( Y_dY_d^\dagger \right)
+ (-4/3)^2 \, {\mathrm{Tr}}\left( Y_uY_u^\dagger \right)
\bigg] 
\nonumber\\ && \qquad
+ \frac{3}{20} \bigg[ (-1)^2 \, {\mathrm{Tr}}\left( Y_eY_e^\dagger \right)
+ (2)^2 \, {\mathrm{Tr}}\left( Y_eY_e^\dagger \right) \bigg]
\Bigg\}_{SM}
\nonumber\\ &&
+ \frac{3}{20}(1)^2 \left[ \left(\frac{\tilde g_u}{\sqrt{2}}\right)^2 \times 3
+ \left(\frac{\tilde g'_u}{\sqrt{2}}\right)^2 \right]
+ \frac{3}{20}(-1)^2 \left[ \left(\frac{\tilde g_d}{\sqrt{2}}\right)^2 \times 3
+ \left(\frac{\tilde g'_d}{\sqrt{2}}\right)^2 \right]
\\ &=&
\Bigg\{ \frac{17}{20} {\mathrm{Tr}}\left( Y_uY_u^\dagger \right) +
\frac{1}{4} {\mathrm{Tr}}\left( Y_dY_d^\dagger \right) +
\frac{3}{4} {\mathrm{Tr}}\left( Y_eY_e^\dagger \right) \Bigg\}_{SM} +
\frac{9}{40}\Big( \tilde g^2_u+\tilde g^2_d \Big) +
\frac{3}{40}\Big( \tilde g'^2_u+\tilde g'^2_d \Big)
\nonumber
\end{eqnarray}
where we have included $Q$, $d_R$, $u_R$, $L$, $e_R$, $\widetilde H_u$,
and $\widetilde H_d$. Notice that the Yukawa terms that contribute due to
the higgsinos are written in eq.~(\ref{HHinoGino}); the term associated to 
$SU(2)$ is,
\begin{eqnarray}
Y_2(F) &=& \Bigg\{
\frac{1}{3} \times \frac{3}{4} \bigg[ 
3 \times {\mathrm{Tr}}\left( Y_uY_u^\dagger + Y_dY_d^\dagger \right) +
{\mathrm{Tr}}\left( Y_eY_e^\dagger \right)
\bigg] \Bigg\}_{SM}
+ \frac{1}{3} \times 2
\left[ \left(\frac{\tilde g_u}{\sqrt{2}}\right)^2
+ \left(\frac{\tilde g_d}{\sqrt{2}}\right)^2 \right] \times 3
\nonumber\\ &&
+ \frac{1}{3} \times \frac{3}{4}
\left[ \left(\frac{\tilde g_u}{\sqrt{2}}\right)^2 \times 3
+ \left(\frac{\tilde g'_u}{\sqrt{2}}\right)^2 \right]
+ \frac{1}{3} \times \frac{3}{4}
\left[ \left(\frac{\tilde g_d}{\sqrt{2}}\right)^2 \times 3
+ \left(\frac{\tilde g'_d}{\sqrt{2}}\right)^2 \right]
\\ &=& \Bigg\{
\frac{3}{4} {\mathrm{Tr}}\left( Y_uY_u^\dagger \right) +
\frac{3}{4} {\mathrm{Tr}}\left( Y_dY_d^\dagger \right) +
\frac{1}{4} {\mathrm{Tr}}\left( Y_eY_e^\dagger \right)
\Bigg\}_{SM} + \frac{11}{8} \left( \tilde g_u^2+\tilde g_d^2 \right)
+ \frac{1}{8} \left( \tilde g'^2_u+\tilde g'^2_d \right)
\nonumber
\end{eqnarray}
where we have included $Q$, $L$, $\widetilde W$, $\widetilde H_u$, and
$\widetilde H_d$; the term associated to $SU(3)$ is,
\begin{eqnarray}
Y_3(F) &=& \Bigg\{
\frac{1}{8} \times \frac{4}{3} \times 3 \bigg[
{\mathrm{Tr}}\left( Y_uY_u^\dagger + Y_dY_d^\dagger \right)
+ {\mathrm{Tr}}\left( Y_dY_d^\dagger \right)
+ {\mathrm{Tr}}\left( Y_uY_u^\dagger \right)
\bigg] \Bigg\}_{SM}
\nonumber\\
&=& \Bigg\{ {\mathrm{Tr}}\left( Y_uY_u^\dagger \right) +
{\mathrm{Tr}}\left( Y_dY_d^\dagger \right) \Bigg\}_{SM}
\end{eqnarray}
where we have included $Q$, $d_R$, and $u_R$.

To summarize, the two-loop RGE for the gauge couplings in PPS are, for $U(1)$
\begin{eqnarray}
\frac{dg_1}{dt} &=& \frac{g_1^3}{16\pi^2} \frac{23}{5} +
\frac{g_1^3}{(16\pi^2)^2} \bigg[
\frac{217}{50}g_1^2+\frac{9}{2}g_2^2+\frac{44}{5}g_3^2 -
\frac{9}{20}\left( \tilde g^2_u+\tilde g^2_d \right) -
\frac{3}{20}\left( \tilde g'^2_u+\tilde g'^2_d \right)
\nonumber\\ && \qquad\qquad\qquad\qquad\,\,
- \frac{17}{10} {\mathrm{Tr}}\left( Y_uY_u^\dagger \right)
- \frac{1}{2} {\mathrm{Tr}}\left( Y_dY_d^\dagger \right)
- \frac{3}{2} {\mathrm{Tr}}\left( Y_eY_e^\dagger \right)
\bigg]
\end{eqnarray}
for $SU(2)$,
\begin{eqnarray}
\frac{dg_2}{dt} &=& -\frac{g_2^3}{16\pi^2} +
\frac{g_2^3}{(16\pi^2)^2} \bigg[
\frac{3}{2}g_1^2+\frac{225}{6}g_2^2+12g_3^2 -
\frac{11}{4}\left( \tilde g^2_u+\tilde g^2_d \right) -
\frac{1}{4}\left( \tilde g'^2_u+\tilde g'^2_d \right)
\nonumber\\ && \qquad\qquad\qquad\qquad\,\,
- \frac{3}{2} {\mathrm{Tr}}\left( Y_uY_u^\dagger \right)
- \frac{3}{2} {\mathrm{Tr}}\left( Y_dY_d^\dagger \right)
- \frac{1}{2} {\mathrm{Tr}}\left( Y_eY_e^\dagger \right)
\bigg]
\end{eqnarray}
and for $SU(3)$,
\begin{equation}
\frac{dg_3}{dt} = -5\frac{g_3^3}{16\pi^2} +
\frac{g_3^3}{(16\pi^2)^2} \bigg[
\frac{11}{10}g_1^2+\frac{9}{2}g_2^2+22g_3^2
- 2 {\mathrm{Tr}}\left( Y_uY_u^\dagger \right)
- 2 {\mathrm{Tr}}\left( Y_dY_d^\dagger \right)
\bigg]
\end{equation}
%

%%%%%%%%%%%%%%%%%%%%%%%%%%%%%%%%%%%%%%%%%%%%%%%%%%%%%%%%%%%%%%%%%%
\subsection{General One-loop RGE for Yukawa Couplings}
%%%%%%%%%%%%%%%%%%%%%%%%%%%%%%%%%%%%%%%%%%%%%%%%%%%%%%%%%%%%%%%%%%

In the context of Renormalization Group Equations, the usual notation 
for the Yukawa terms in the lagrangian is,
\begin{equation}
{\cal L}_Y = -\psi^\dagger_i Y_{ij}^a \psi_j \phi_a
-\psi^\dagger_j Y_{ij}^{a*} \psi_i \phi_a
\end{equation}
where $Y^a$ is a hermitic matrix, $\psi_i$ $i=1...n$ are $n$ two-component
fermion fields in chiral representation, and $\phi_a$ $i=1...m$ are $m$
real scalar fields. The $m$ RGEs for the $n\times n$ Yukawa matrices are
\begin{eqnarray}
(16\pi^2) \frac{d}{dt} {\bf Y}^a &=&
\frac{1}{2} {\bf Y}^{b\dagger} {\bf Y}^b {\bf Y}^a +
\frac{1}{2} {\bf Y}^a {\bf Y}^{b\dagger} {\bf Y}^b +
2 {\bf Y}^b {\bf Y}^{a\dagger} {\bf Y}^b
\nonumber\\ &+&
\frac{1}{2} {\bf Y}^b \, {\mathrm{Tr}}
\left( {\bf Y}^{b\dagger} {\bf Y}^a+{\bf Y}^{a\dagger} {\bf Y}^b \right)
- C^a(Y^a_{ij}) {\bf Y}^a
\label{YaRGE}
\end{eqnarray}
with
\begin{equation}
C^a(Y^a_{ij}) = 3\sum_{f,k} g_k^2 C_2(R_k)
\end{equation}
In this last equation the sum over $f$ runs over all fermions that couple by 
${\bf Y}^a_{ij}$, $C_2(R_k)$ is the normalization of the Casimir operator 
in the representation $R_k$ of the gauge groups with coupling constant $g_k$, 
$k=1,2,3$.

%%%%%%%%%%%%%%%%%%%%%%%%%%%%%%%%%%%%%%%%%%%%%%%%%%%%%%%%%%%%%%%%%%
\subsection{Explicit One-loop RGE for Yukawa Couplings in PSS}
%%%%%%%%%%%%%%%%%%%%%%%%%%%%%%%%%%%%%%%%%%%%%%%%%%%%%%%%%%%%%%%%%%

In our model the terms in the lagrangian usually known as Yukawa couplings are
\begin{equation}
{\cal L}_Y = \overline u_R {\bf h}_u H_u^T \varepsilon \, Q_L -
\overline d_R {\bf h}_d H_d^T \varepsilon \, Q_L -
\overline e_R {\bf h}_e H_d^T \varepsilon L_L
\label{Yuk}
\end{equation}
The couplings themselves, ${\bf h}_u$, ${\bf h}_d$, and ${\bf h}_e$, are $3\times3$
matrices in flavour space. The first step is to decompose the Higgs doublets into
their real scalar components
\begin{equation}
H_u = \frac{1}{\sqrt{2}} \left( 
\begin{matrix} \phi^u_1+i\phi^u_2 \cr \phi^u_3+i\phi^u_4 \end{matrix} \right)
\,,\qquad
H_d = \frac{1}{\sqrt{2}} \left( 
\begin{matrix} \phi^d_1+i\phi^d_2 \cr \phi^d_3+i\phi^d_4 \end{matrix} \right)
\end{equation}
such that the terms in eq.~(\ref{Yuk}) become
\begin{eqnarray}
{\cal L}_Y &=&
\frac{1}{\sqrt{2}} \overline u_R {\bf h}_u \Big( \phi^u_1+i\phi^u_2 \Big) d_L -
\frac{1}{\sqrt{2}} \overline u_R {\bf h}_u \Big( \phi^u_3+i\phi^u_4 \Big) u_L -
\frac{1}{\sqrt{2}} \overline d_R {\bf h}_d \Big( \phi^d_1+i\phi^d_2 \Big) d_L
\nonumber\\ && +
\frac{1}{\sqrt{2}} \overline d_R {\bf h}_d \Big( \phi^d_3+i\phi^d_4 \Big) u_L -
\frac{1}{\sqrt{2}} \overline e_R {\bf h}_e \Big( \phi^d_1+i\phi^d_2 \Big) e_L +
\frac{1}{\sqrt{2}} \overline e_R {\bf h}_e \Big( \phi^d_3+i\phi^d_4 \Big) \nu_L
\label{YukawaExp}
\end{eqnarray}
To these terms we have to add the Higgs-higgsino-gaugino terms in 
eq.~(\ref{HHinoGino}), which also are Yukawa type terms. After the 
decomposition we write them as,
\begin{eqnarray}
{\cal L}_{Hhg} &=& - \frac{1}{2} \Big( \phi^u_1-i\phi^u_2 \Big)
\bigg[ \Big( \tilde g_u \widetilde W^3 + \tilde g'_u \widetilde B \Big)
\widetilde H^+_u + \tilde g_u \Big( \widetilde W^1-i\widetilde W^2 \Big)
\widetilde H^0_u \bigg]
\nonumber\\ &&
- \frac{1}{2} \Big( \phi^u_3-i\phi^u_4 \Big)
\bigg[ \Big( -\tilde g_u \widetilde W^3 + \tilde g'_u \widetilde B \Big)
\widetilde H^0_u + \tilde g_u \Big( \widetilde W^1+i\widetilde W^2 \Big)
\widetilde H^+_u \bigg]
\nonumber\\ &&
- \frac{1}{2} \Big( \phi^d_1-i\phi^d_2 \Big)
\bigg[ \Big( \tilde g_d \widetilde W^3 - \tilde g'_d \widetilde B \Big)
\widetilde H^0_d + \tilde g_d \Big( \widetilde W^1-i\widetilde W^2 \Big)
\widetilde H^-_d \bigg]
\label{HhgExpand}\\ &&
- \frac{1}{2} \Big( \phi^d_3-i\phi^d_4 \Big)
\bigg[ - \Big( \tilde g_d \widetilde W^3 + \tilde g'_d \widetilde B \Big)
\widetilde H^-_d + \tilde g_d \Big( \widetilde W^1+i\widetilde W^2 \Big)
\widetilde H^0_d \bigg]
\nonumber
\end{eqnarray}
The terms in eqs.~(\ref{YukawaExp}) and (\ref{HhgExpand}) are grouped into
eight matrices ${\bf Y}^a$, whose index $a$ runs over the eight real scalar 
fields we have, namely, $a=\phi^u_1$, $\phi^u_2$, $\phi^u_3$, $\phi^u_4$, 
$\phi^d_1$, $\phi^d_2$, $\phi^d_3$, $\phi^d_4$. Each of these matrices is 
$15\times 15$, and it is expanded in the base formed by the 15 fermions 
$\nu_L$, $e_L$, $e_R$, $u_L$, $d_L$, $u_R$, $d_R$, $\widetilde W^1$, 
$\widetilde W^2$, $\widetilde W^3$, $\widetilde B$, $\widetilde H^+_u$, 
$\widetilde H^0_u$, $\widetilde H^0_d$, $\widetilde H^-_d$. These are the 
matrices whose RGE are given in eq.~(\ref{YaRGE}).

Within these RGE, we see the coefficients $C^a(Y^a_{ij})$, which we calculate now.
The four scalars belonging to $H_u$ have the same coefficient, which associated to
the Yukawa coupling ${\bf h}_u$ is,
\begin{equation}
C^{\phi^u_i}({\bf h}_u) = \Bigg\{
3g_1^2 \times \frac{3}{20} \Big[ (1/3)^2 + (-4/3)^2 \Big] +
3g_2^2 \times \frac{3}{4} + 3g_3^2 \Big[ \frac{4}{3}+\frac{4}{3} \Big]
\Bigg\}_{SM} =
\frac{17}{20} g_1^2 + \frac{9}{4} g_2^2 + 8 g_3^2
\end{equation}
where we have included $Q$ and $u_R$ for $U(1)$ and $SU(3)$, and only $Q$ for
$SU(2)$. The same coefficient associated to the Yukawa coupling $\tilde g_u$ is,
\begin{equation}
C^{\phi^u_i}(\tilde g_u) = \Bigg\{ 3g_1^2 \times \frac{3}{20} (1)^2 +
3g_2^2 \Big[ 2 + \frac{3}{4} \Big] \Bigg\}_{SM} =
\frac{9}{20} g_1^2 + \frac{33}{4} g_2^2
\end{equation}
where we included $\widetilde H_u$ for $U(1)$, $\widetilde W$ and $\widetilde H_u$
for $SU(2)$, and none for $SU(3)$. Lastly, the same coefficient but this time 
associated to the Yukawa coupling $\tilde g'_u$ is,
\begin{equation}
C^{\phi^u_i}(\tilde g'_u) = \Bigg\{ 3g_1^2 \times \frac{3}{20} (1)^2 +
3g_2^2 \times \frac{3}{4} \Bigg\}_{SM} =
\frac{9}{20} g_1^2 + \frac{9}{4} g_2^2
\end{equation}
where we have included $\widetilde H_u$ for $U(1)$ and $SU(2)$, and no fields for 
$SU(3)$.

This allow us to calculate the RGEs for couplings associated to the 
entries of the matrix ${\bf Y}^{\phi^u_1}$,
\begin{eqnarray}
(16\pi^2) \frac{d}{dt} {\bf h}_u &=& {\bf h}_u \left\{
\frac{3}{2}\tilde g_u^2 + \frac{1}{2}\tilde g'^2_u +
\frac{3}{2}{\bf h}_u^\dagger {\bf h}_u+\frac{1}{2}{\bf h}_d^\dagger {\bf h}_d
+ 3 {\mathrm{Tr}} \left( {\bf h}_u^\dagger {\bf h}_u \right) -
\frac{17}{20} g_1^2 - \frac{9}{4} g_2^2 - 8 g_3^2 \right\}
\nonumber\\
(16\pi^2) \frac{d}{dt} \tilde g_u &=& \tilde g_u \left\{
\frac{11}{4} \tilde g^2_u + \frac{3}{4}\tilde g'^2_u +\frac{1}{2}\tilde g^2_d +
3 {\mathrm{Tr}} \left( {\bf h}_u^\dagger {\bf h}_u \right) -
\frac{9}{20}g_1^2-\frac{33}{4}g_2^2 \right\}
\label{uRGE}\\
(16\pi^2) \frac{d}{dt} \tilde g'_u &=& \tilde g'_u \left\{
\frac{9}{4}\tilde g^2_u + \frac{5}{4}\tilde g'^2_u + \frac{1}{2}\tilde g'^2_d
+ 3 {\mathrm{Tr}} \left( {\bf h}_u^\dagger {\bf h}_u \right)
- \frac{9}{20}g_1^2 - \frac{9}{4}g_2^2 \right\}
\nonumber
\end{eqnarray}

Now we continue with the coefficients $C^a(Y^a_{ij})$. As before, the four scalars
belonging to $H_d$ have the same coefficient. Associated to the Yukawa coupling 
${\bf h}_e$ we have,
\begin{equation}
C^{\phi^d_i}({\bf h}_e) = \Bigg\{
3g_1^2 \times \frac{3}{20} \Big[ (-1)^2 + (2)^2 \Big] +
3g_2^2 \times \frac{3}{4}
\Bigg\}_{SM} =
\frac{9}{4} g_1^2 + \frac{9}{4} g_2^2
\end{equation}
where we have included $L$ and $e_R$ for $U(1)$, $L$ for $SU(2)$, and
none for $SU(3)$. Similarly, the coefficient associated to ${\bf h}_d$ is,
\begin{equation}
C^{\phi^d_i}({\bf h}_d) = \Bigg\{
3g_1^2 \times \frac{3}{20} \Big[ (1/3)^2 + (2/3)^2 \Big] +
3g_2^2 \times \frac{3}{4} + 3g_3^2 \Big[ \frac{4}{3}+\frac{4}{3} \Big]
\Bigg\}_{SM} =
\frac{1}{4} g_1^2 + \frac{9}{4} g_2^2 + 8 g_3^2
\end{equation}
where we have included $Q$ and $d_R$ for $U(1)$ and $SU(3)$, and only $Q$ for
$SU(2)$. Finally, note that
\begin{equation}
C^{\phi^u_i}(\tilde g_u)=C^{\phi^d_i}(\tilde g_d)
\,,\qquad
C^{\phi^u_i}(\tilde g'_u)=C^{\phi^d_i}(\tilde g'_d)
\end{equation}

With these results the four missing RGE from the entries of matrix
${\bf Y}^{\phi^d_1}$ are,
\begin{eqnarray}
(16\pi^2) \frac{d}{dt} {\bf h}_e &=&
{\bf h}_e \left\{
\frac{3}{2}\tilde g_d^2 + \frac{1}{2}\tilde g'^2_d +
\frac{3}{2} {\bf h}_e^\dagger {\bf h}_e
+ 3 {\mathrm{Tr}} \left( {\bf h}_d^\dagger {\bf h}_d \right) +
{\mathrm{Tr}} \left( {\bf h}_e^\dagger {\bf h}_e \right)
- \frac{9}{4} g_1^2 - \frac{9}{4} g_2^2 \right\}
\nonumber \\
(16\pi^2) \frac{d}{dt} {\bf h}_d &=&
{\bf h}_d \left\{
\frac{3}{2}\tilde g_d^2 + \frac{1}{2}\tilde g'^2_d +
\frac{1}{2} {\bf h}_u^\dagger {\bf h}_u +
\frac{3}{2} {\bf h}_d^\dagger {\bf h}_d +
3 {\mathrm{Tr}} \left( {\bf h}_d^\dagger {\bf h}_d \right) +
{\mathrm{Tr}} \left( {\bf h}_e^\dagger {\bf h}_e \right)
- \frac{1}{4} g_1^2 - \frac{9}{4} g_2^2 - 8 g_3^2 \right\}
\nonumber \\
(16\pi^2) \frac{d}{dt} \tilde g_d &=&
\tilde g_d \left\{
\frac{1}{2} \tilde g^2_u +
\frac{11}{4}\tilde g^2_d+\frac{3}{4}\tilde g'^2_d
+ 3 {\mathrm{Tr}} \left( {\bf h}_d^\dagger {\bf h}_d \right) +
{\mathrm{Tr}} \left( {\bf h}_e^\dagger {\bf h}_e \right)
- \frac{9}{20} g_1^2 - \frac{33}{4} g_2^2 \right\}
\label{dRGE}\\
(16\pi^2) \frac{d}{dt} \tilde g'_d &=&
\tilde g'_d \left\{
\frac{1}{2} \tilde g'^2_u +
\frac{9}{4}\tilde g^2_d+\frac{5}{4}\tilde g'^2_d
+ 3 {\mathrm{Tr}} \left( {\bf h}_d^\dagger {\bf h}_d \right) +
{\mathrm{Tr}} \left( {\bf h}_e^\dagger {\bf h}_e \right)
- \frac{9}{20} g_1^2 - \frac{9}{4} g_2^2 \right\}
\nonumber
\end{eqnarray}
Thus, eqs.~(\ref{uRGE}) and (\ref{dRGE}) are the RGE for the Yukawa couplings 
for the Partial Split Supersymmetry Model.

\newpage

%%%%%%%%%%%%%%%%%%%%%%%%%%%%%%%%%%%%%%%%%%%%%%%%%%%%%%%%%%%%%%%%%%%%%%%%%%%%%%


\begin{thebibliography}{99}
%%%%%%%%%%%%%%%%%%%%%%%%%%%%%%%%%%%%%%%%%%%%%%%%%%%%%%%%%%%%%%%%%%%%%%%%%%%%%%

%\cite{:2012gk}
\bibitem{:2012gk} 
  G.~Aad {\it et al.}  [ATLAS Collaboration],
  %``Observation of a new particle in the search for the Standard Model Higgs boson with the ATLAS detector at the LHC,''
  Phys.\ Lett.\ B {\bf 716}, 1 (2012)
  [arXiv:1207.7214 [hep-ex]].
  %%CITATION = ARXIV:1207.7214;%%

%\cite{:2012gu}
\bibitem{:2012gu} 
  S.~Chatrchyan {\it et al.}  [CMS Collaboration],
  %``Observation of a new boson at a mass of 125 GeV with the CMS experiment at the LHC,''
  Phys.\ Lett.\ B {\bf 716}, 30 (2012)
  [arXiv:1207.7235 [hep-ex]].
  %%CITATION = ARXIV:1207.7235;%%


%\cite{Aad:2011cwa}
\bibitem{Aad:2011cwa} 
  G.~Aad {\it et al.}  [ATLAS Collaboration],
  %``Searches for supersymmetry with the ATLAS detector using final states with two leptons and missing transverse momentum in sqrt{s} = 7 TeV proton-proton collisions,''
  arXiv:1110.6189 [hep-ex].
  %%CITATION = ARXIV:1110.6189;%%

%\cite{Aad:2011zj}
\bibitem{Aad:2011zj} 
  G.~Aad {\it et al.}  [ATLAS Collaboration],
  %``Search for Diphoton Events with Large Missing Transverse Momentum in 1 fb^-1 of 7 TeV Proton-Proton Collision Data with the ATLAS Detector,''
  arXiv:1111.4116 [hep-ex].
  %%CITATION = ARXIV:1111.4116;%%

%\cite{Aad:2011cw}
\bibitem{Aad:2011cw} 
  [ATLAS Collaboration],
  %``Search for scalar bottom pair production with the ATLAS detector in pp Collisions at sqrt{s} = 7 TeV,''
  arXiv:1112.3832 [hep-ex].
  %%CITATION = ARXIV:1112.3832;%%

%\cite{Aad:2012zn}
\bibitem{Aad:2012zn} 
  G.~Aad {\it et al.}  [ATLAS Collaboration],
  %``Search for decays of stopped, long-lived particles from 7 TeV pp collisions with the ATLAS detector,''
  arXiv:1201.5595 [hep-ex].
  %%CITATION = ARXIV:1201.5595;%%
%\cite{Aad:2011yh}
%\bibitem{Aad:2011yh} 
  G.~Aad {\it et al.}  [ATLAS Collaboration],
  %``Search for Massive Colored Scalars in Four-Jet Final States in sqrt{s}=7 TeV proton-proton collisions with the ATLAS Detector,''
  Eur.\ Phys.\ J.\ C {\bf 71}, 1828 (2011)
  [arXiv:1110.2693 [hep-ex]].
  %%CITATION = ARXIV:1110.2693;%%
%\cite{Aad:2011qa}
%\bibitem{Aad:2011qa} 
  G.~Aad {\it et al.}  [Atlas Collaboration],
  %``Search for new phenomena in final states with large jet multiplicities and missing transverse momentum using sqrt(s)=7 TeV pp collisions with the ATLAS detector,''
  JHEP {\bf 1111}, 099 (2011)
  [arXiv:1110.2299 [hep-ex]].
  %%CITATION = ARXIV:1110.2299;%%
%\cite{Aad:2011ib}
%\bibitem{Aad:2011ib} 
  G.~Aad {\it et al.}  [ATLAS Collaboration],
  %``Search for squarks and gluinos using final states with jets and missing transverse momentum with the ATLAS detector in sqrt(s) = 7 TeV proton-proton collisions,''
  arXiv:1109.6572 [hep-ex].
  %%CITATION = ARXIV:1109.6572;%%
%\cite{Aad:2011kz}
%\bibitem{Aad:2011kz} 
  G.~Aad {\it et al.}  [ATLAS Collaboration],
  %``Search for Diphoton Events with Large Missing Transverse Energy with 36 pb^-1 of 7 TeV Proton-Proton Collision Data with the ATLAS Detector,''
  Eur.\ Phys.\ J.\ C {\bf 71}, 1744 (2011)
  [arXiv:1107.0561 [hep-ex]].
  %%CITATION = ARXIV:1107.0561;%%
%\cite{Aad:2011hz}
%\bibitem{Aad:2011hz} 
  G.~Aad {\it et al.}  [ATLAS Collaboration],
  %``Search for Heavy Long-Lived Charged Particles with the ATLAS detector in pp collisions at sqrt(s) = 7 TeV,''
  Phys.\ Lett.\ B {\bf 703}, 428 (2011)
  [arXiv:1106.4495 [hep-ex]].
  %%CITATION = ARXIV:1106.4495;%%
%\cite{Aad:2011xk}
%\bibitem{Aad:2011xk} 
  G.~Aad {\it et al.}  [ATLAS Collaboration],
  %``Search for an excess of events with an identical flavour lepton pair and significant missing transverse momentum in sqrt{s} = 7 TeV proton-proton collisions with the ATLAS detector,''
  Eur.\ Phys.\ J.\ C {\bf 71}, 1647 (2011)
  [arXiv:1103.6208 [hep-ex]].
  %%CITATION = ARXIV:1103.6208;%%
%\cite{Aad:2011xm}
%\bibitem{Aad:2011xm} 
  G.~Aad {\it et al.}  [ATLAS Collaboration],
  %``Search for supersymmetric particles in events with lepton pairs and large missing transverse momentum in $\sqrt{s}=7$ TeV proton-proton collisions with the ATLAS experiment,''
  Eur.\ Phys.\ J.\ C {\bf 71}, 1682 (2011)
  [arXiv:1103.6214 [hep-ex]].
  %%CITATION = ARXIV:1103.6214;%%
%\cite{Aad:2011kta}
%\bibitem{Aad:2011kta} 
  G.~Aad {\it et al.}  [ATLAS Collaboration],
  %``Search for a heavy particle decaying into an electron and a muon with the ATLAS detector in $\sqrt{s}=7$ TeV $pp$ collisions at the LHC,''
  Phys.\ Rev.\ Lett.\  {\bf 106}, 251801 (2011)
  [arXiv:1103.5559 [hep-ex]].
  %%CITATION = ARXIV:1103.5559;%%
%\cite{Aad:2011ks}
%\bibitem{Aad:2011ks} 
  G.~Aad {\it et al.}  [ATLAS Collaboration],
  %``Search for supersymmetry in pp collisions at sqrt{s} = 7TeV in final states with missing transverse momentum and b-jets,''
  Phys.\ Lett.\ B {\bf 701}, 398 (2011)
  [arXiv:1103.4344 [hep-ex]].
  %%CITATION = ARXIV:1103.4344;%%
%\cite{Aad:2011yf}
%\bibitem{Aad:2011yf} 
  G.~Aad {\it et al.}  [ATLAS Collaboration],
  %``Search for stable hadronising squarks and gluinos with the ATLAS experiment at the LHC,''
  Phys.\ Lett.\ B {\bf 701}, 1 (2011)
  [arXiv:1103.1984 [hep-ex]].
  %%CITATION = ARXIV:1103.1984;%%
%\cite{daCosta:2011qk}
%\bibitem{daCosta:2011qk} 
  G.~Aad {\it et al.}  [Atlas Collaboration],
  %``Search for squarks and gluinos using final states with jets and missing transverse momentum with the ATLAS detector in sqrt(s) = 7 TeV proton-proton collisions,''
  Phys.\ Lett.\ B {\bf 701}, 186 (2011)
  [arXiv:1102.5290 [hep-ex]].
  %%CITATION = ARXIV:1102.5290;%%
%\cite{Aad:2011hh}
%\bibitem{Aad:2011hh} 
  G.~Aad {\it et al.}  [Atlas Collaboration],
  %``Search for supersymmetry using final states with one lepton, jets, and missing transverse momentum with the ATLAS detector in sqrt{s} = 7 TeV pp,''
  Phys.\ Rev.\ Lett.\  {\bf 106}, 131802 (2011)
  [arXiv:1102.2357 [hep-ex]].
  %%CITATION = ARXIV:1102.2357;%%

%\cite{Chatrchyan:2011wc}
\bibitem{Chatrchyan:2011wc} 
  S.~Chatrchyan {\it et al.}  [CMS Collaboration],
  %``Search for Supersymmetry in $pp$ Collisions at $\sqrt{s} = 7$ TeV in Events with Two Photons and Missing Transverse Energy,''
  Phys.\ Rev.\ Lett.\  {\bf 106}, 211802 (2011)
  [arXiv:1103.0953 [hep-ex]].
  %%CITATION = ARXIV:1103.0953;%%
%\cite{Chatrchyan:2011bz}
%\bibitem{Chatrchyan:2011bz} 
  S.~Chatrchyan {\it et al.}  [CMS Collaboration],
  %``Search for Physics Beyond the Standard Model in Opposite-Sign Dilepton Events at $\sqrt{s} = 7$ TeV,''
  JHEP {\bf 1106}, 026 (2011)
  [arXiv:1103.1348 [hep-ex]].
  %%CITATION = ARXIV:1103.1348;%%
%\cite{Chatrchyan:2011wba}
%\bibitem{Chatrchyan:2011wba} 
  S.~Chatrchyan {\it et al.}  [CMS Collaboration],
  %``Search for new physics with same-sign isolated dilepton events with jets and missing transverse energy at the LHC,''
  JHEP {\bf 1106}, 077 (2011)
  [arXiv:1104.3168 [hep-ex]].
  %%CITATION = ARXIV:1104.3168;%%
%\cite{Chatrchyan:2011ah}
%\bibitem{Chatrchyan:2011ah} 
  S.~Chatrchyan {\it et al.}  [CMS Collaboration],
  %``Search for supersymmetry in events with a lepton, a photon, and large missing transverse energy in pp collisions at sqrt(s) = 7 TeV,''
  JHEP {\bf 1106}, 093 (2011)
  [arXiv:1105.3152 [hep-ex]].
  %%CITATION = ARXIV:1105.3152;%%
%\cite{Chatrchyan:2011ff}
%\bibitem{Chatrchyan:2011ff} 
  S.~Chatrchyan {\it et al.}  [CMS Collaboration],
  %``Search for Physics Beyond the Standard Model Using Multilepton Signatures in pp Collisions at sqrt(s)=7 TeV,''
  Phys.\ Lett.\ B {\bf 704}, 411 (2011)
  [arXiv:1106.0933 [hep-ex]].
  %%CITATION = ARXIV:1106.0933;%%
%\cite{Chatrchyan:2011bj}
%\bibitem{Chatrchyan:2011bj} 
  S.~Chatrchyan {\it et al.}  [CMS Collaboration],
  %``Search for Supersymmetry in Events with b Jets and Missing Transverse Momentum at the LHC,''
  JHEP {\bf 1107}, 113 (2011)
  [arXiv:1106.3272 [hep-ex]].
  %%CITATION = ARXIV:1106.3272;%%
%\cite{Collaboration:2011ida}
%\bibitem{Collaboration:2011ida} 
  S.~Chatrchyan {\it et al.}  [CMS Collaboration],
  %``Search for New Physics with Jets and Missing Transverse Momentum in pp collisions at sqrt(s) = 7 TeV,''
  JHEP {\bf 1108}, 155 (2011)
  [arXiv:1106.4503 [hep-ex]].
  %%CITATION = ARXIV:1106.4503;%%
%\cite{Chatrchyan:2011ek}
%\bibitem{Chatrchyan:2011ek} 
  S.~Chatrchyan {\it et al.}  [CMS Collaboration],
  %``Inclusive search for squarks and gluinos in pp collisions at sqrt(s) = 7 TeV,''
  Phys.\ Rev.\ D {\bf 85}, 012004 (2012)
  [arXiv:1107.1279 [hep-ex]].
  %%CITATION = ARXIV:1107.1279;%%
%\cite{Chatrchyan:2011qs}
%\bibitem{Chatrchyan:2011qs} 
  S.~Chatrchyan {\it et al.}  [CMS Collaboration],
  %``Search for supersymmetry in pp collisions at sqrt(s)=7 TeV in events with a single lepton, jets, and missing transverse momentum,''
  JHEP {\bf 1108}, 156 (2011)
  [arXiv:1107.1870 [hep-ex]].
  %%CITATION = ARXIV:1107.1870;%%
%\cite{Chatrchyan:2011zy}
%\bibitem{Chatrchyan:2011zy} 
  S.~Chatrchyan {\it et al.}  [CMS Collaboration],
  %``Search for Supersymmetry at the LHC in Events with Jets and Missing Transverse Energy,''
  Phys.\ Rev.\ Lett.\  {\bf 107}, 221804 (2011)
  [arXiv:1109.2352 [hep-ex]].
  %%CITATION = ARXIV:1109.2352;%%

%\cite{Aad:2011zb}
\bibitem{Aad:2011zb} 
  G.~Aad {\it et al.}  [ATLAS Collaboration],
  %``Search for displaced vertices arising from decays of new heavy particles in 7 TeV pp collisions at ATLAS,''
  Phys.\ Lett.\ B {\bf 707}, 478 (2012)
  [arXiv:1109.2242 [hep-ex]].
  %%CITATION = ARXIV:1109.2242;%%

%\cite{Collaboration:2011qr}
\bibitem{Collaboration:2011qr} 
  A.~Collaboration,
  %``Search for a heavy neutral particle decaying into an electron and a muon using 1 fb^-1 of ATLAS data,''
  Eur.\ Phys.\ J.\ C {\bf 71}, 1809 (2011)
  [arXiv:1109.3089 [hep-ex]].
  %%CITATION = ARXIV:1109.3089;%%

%\cite{ATLAS:2011ad}
\bibitem{ATLAS:2011ad} 
  G.~Aad {\it et al.}  [ATLAS Collaboration],
  %``Search for supersymmetry in final states with jets, missing transverse momentum and one isolated lepton in sqrt{s} = 7 TeV pp collisions using 1 $fb^{-1}$ of ATLAS data,''
  Phys.\ Rev.\ D {\bf 85}, 012006 (2012)
  [arXiv:1109.6606 [hep-ex]].
  %%CITATION = ARXIV:1109.6606;%%

%\cite{ArkaniHamed:2004fb}
\bibitem{ArkaniHamed:2004fb}
  N.~Arkani-Hamed and S.~Dimopoulos,
  %``Supersymmetric unification without low energy supersymmetry and  signatures
  %for fine-tuning at the LHC,''
  JHEP {\bf 0506}, 073 (2005)
  [arXiv:hep-th/0405159].
  %%CITATION = JHEPA,0506,073;%%

%\cite{Giudice:2004tc}
\bibitem{Giudice:2004tc}
  G.~F.~Giudice and A.~Romanino,
  %``Split supersymmetry,''
  Nucl.\ Phys.\  B {\bf 699}, 65 (2004)
  [Erratum-ibid.\  B {\bf 706}, 65 (2005)]
  [arXiv:hep-ph/0406088].
  %%CITATION = NUPHA,B699,65;%%

%\cite{Berezinsky:2004zb}
\bibitem{Berezinsky:2004zb}
  V.~Berezinsky, M.~Narayan and F.~Vissani,
  %``Low scale gravity as the source of neutrino masses?,''
  JHEP {\bf 0504}, 009 (2005)
  [arXiv:hep-ph/0401029].
  %%CITATION = JHEPA,0504,009;%%
%\cite{Diaz:2009yz}
%\bibitem{Diaz:2009yz}
  M.~A.~Diaz, B.~Koch and B.~Panes,
  %``Gravity Effects on Neutrino Masses in Split Supersymmetry,''
  Phys.\ Rev.\  D {\bf 79}, 113009 (2009)
  [arXiv:0902.1720 [hep-ph]].
  %%CITATION = PHRVA,D79,113009;%%

%\cite{Diaz:2006ee}
\bibitem{Diaz:2006ee} 
  M.~A.~Diaz, P.~Fileviez Perez and C.~Mora,
  %``Neutrino Masses in Split Supersymmetry,''
  Phys.\ Rev.\ D {\bf 79}, 013005 (2009)
  [hep-ph/0605285].
  %%CITATION = HEP-PH/0605285;%%

%\cite{Diaz:2009gf}
\bibitem{Diaz:2009gf}
  M.~A.~Diaz, F.~Garay and B.~Koch,
  %``Explaining Solar Neutrinos with Heavy Higgs Masses in Partial Split
  %Supersymmetry,''
  Phys.\ Rev.\  D {\bf 80}, 113005 (2009)
  [arXiv:0910.2987 [hep-ph]].
  %%CITATION = PHRVA,D80,113005;%%

%\cite{Cottin:2011fy}
\bibitem{Cottin:2011fy}
  G.~Cottin, M.~A.~Diaz and B.~Koch,
  %``Non-diagonal Charged Lepton Yukawa Matrix: Effects on Neutrino Mixing in
  %Supersymmetry,''
  arXiv:1112.6351 [hep-ph].
  %%CITATION = ARXIV:1112.6351;%%

%\cite{Roszkowski:2004jd}
\bibitem{Roszkowski:2004jd}
  L.~Roszkowski, R.~Ruiz de Austri and K.~Y.~Choi,
  %``Gravitino dark matter in the CMSSM and implications for leptogenesis  and
  %the LHC,''
  JHEP {\bf 0508}, 080 (2005)
  [arXiv:hep-ph/0408227].
  %%CITATION = JHEPA,0508,080;%%
      %\cite{Covi:2008jy}
%      \bibitem{Covi:2008jy}
        L.~Covi, M.~Grefe, A.~Ibarra and D.~Tran,
        %``Unstable Gravitino Dark Matter and Neutrino Flux,''
        JCAP {\bf 0901}, 029 (2009)
        [arXiv:0809.5030 [hep-ph]].
        %%CITATION = JCAPA,0901,029;%%
%\cite{Grefe:2008zz}
%\bibitem{Grefe:2008zz}
  M.~Grefe,
  ``Neutrino signals from gravitino dark matter with broken R-parity,''
DESY-THESIS-2008-043.
  %%CITATION = DESY-THESIS-2008-043;%%
%\cite{Restrepo:2011rj}
%\bibitem{Restrepo:2011rj}
  D.~Restrepo, M.~Taoso, J.~W.~F.~Valle and O.~Zapata,
  %``Gravitino dark matter and neutrino masses with bilinear R-parity
  %violation,''
  Phys.\ Rev.\  D {\bf 85}, 023523 (2012)
  [arXiv:1109.0512 [hep-ph]].
  %%CITATION = PHRVA,D85,023523;%%

%\cite{Diaz:2011pc}
\bibitem{Diaz:2011pc}
  M.~A.~Diaz, S.~G.~Saenz and B.~Koch,
  %``Gravitino Dark Matter and Neutrino Masses in Partial Split Supersymmetry,''
  Phys.\ Rev.\  D {\bf 84}, 055007 (2011)
  [arXiv:1106.0308 [hep-ph]].
  %%CITATION = PHRVA,D84,055007;%%

%\cite{Hirsch:2000ef}
\bibitem{Hirsch:2000ef}
  M.~Hirsch, M.~A.~Diaz, W.~Porod, J.~C.~Romao and J.~W.~F.~Valle,
  %``Neutrino masses and mixings from supersymmetry with bilinear R-parity
  %violation: A theory for solar and atmospheric neutrino oscillations,''
  Phys.\ Rev.\  D {\bf 62}, 113008 (2000)
  [Erratum-ibid.\  D {\bf 65}, 119901 (2002)]
  [arXiv:hep-ph/0004115].
  %%CITATION = PHRVA,D62,113008;%%

%\cite{Hirsch:1997vz}
\bibitem{Hirsch:1997vz}
  M.~Hirsch, H.~V.~Klapdor-Kleingrothaus and S.~G.~Kovalenko,
  %``B-L violating masses in softly broken supersymmetry,''
  Phys.\ Lett.\  B {\bf 398}, 311 (1997)
  [arXiv:hep-ph/9701253].
  %%CITATION = PHLTA,B398,311;%%







%\cite{Hoecker:2010qn}
\bibitem{Hoecker:2010qn}
  A.~Hoecker,
  %``The Hadronic Contribution to the Muon Anomalous Magnetic Moment and to the
  %Running Electromagnetic Fine Structure Constant at MZ - Overview and Latest
  %Results,''
  arXiv:1012.0055 [hep-ph].
  %%CITATION = ARXIV:1012.0055;%%

%\cite{Nakamura:2010zzi}
\bibitem{Nakamura:2010zzi}
  K.~Nakamura {\it et al.}  [Particle Data Group],
  %``Review of particle physics,''
  J.\ Phys.\ G {\bf 37}, 075021 (2010).
  %%CITATION = JPHGB,G37,075021;%%

%\cite{Maltoni:2004ei}
\bibitem{Maltoni:2004ei}
  M.~Maltoni, T.~Schwetz, M.~A.~Tortola, J.~W.~F.~Valle,
  %``Status of global fits to neutrino oscillations,''
  New J.\ Phys.\  {\bf 6}, 122 (2004).
  [hep-ph/0405172].

%\cite{Machacek:1983tz}
\bibitem{Machacek:1983tz}
  M.~E.~Machacek, M.~T.~Vaughn,
  %``Two Loop Renormalization Group Equations in a General Quantum Field Theory. 1. Wave Function Renormalization,''
  Nucl.\ Phys.\  {\bf B222}, 83 (1983);
%\cite{Machacek:1983fi}
%\bibitem{Machacek:1983fi}
  M.~E.~Machacek, M.~T.~Vaughn,
  %``Two Loop Renormalization Group Equations in a General Quantum Field Theory. 2. Yukawa Couplings,''
  Nucl.\ Phys.\  {\bf B236}, 221 (1984);
%\cite{Machacek:1984zw}
%\bibitem{Machacek:1984zw}
  M.~E.~Machacek, M.~T.~Vaughn,
  %``Two Loop Renormalization Group Equations in a General Quantum Field Theory. 3. Scalar Quartic Couplings,''
  Nucl.\ Phys.\  {\bf B249}, 70 (1985).



\end{thebibliography}
\end{document}